\documentclass[10pt,journal]{IEEEtran}
\usepackage{subfigure}
\usepackage{graphicx}
\usepackage{mathrsfs}
\usepackage{epsfig}
\usepackage{times}
\usepackage{amsmath}
\usepackage{float}
\usepackage{amssymb}
\usepackage{txfonts}
\usepackage{stmaryrd}
\usepackage[xetex]{epstopdf}
\usepackage{amsfonts}
\usepackage{newtxtext,newtxmath}
\usepackage{cite}
\usepackage{color}
\usepackage{bm}
\usepackage{multirow}
\usepackage{footnote}
\usepackage[colorlinks,linkcolor=blue,anchorcolor=black,citecolor=blue]{hyperref}

\newtheorem{remark}{Remark}
\newtheorem{theorem}{Theorem}

\begin{document}

\title{
Modeling and Control for UAV with Off-center Slung Load}

\author{Zongyang Lv,~\IEEEmembership{Member,~IEEE,}~
Yanmei Jia,~
Yongqing Liu,~
Alan F. Lynch,~\IEEEmembership{Member,~IEEE,}
Qing Zhao,~\IEEEmembership{Member,~IEEE,}
Yuhu Wu,~\IEEEmembership{Member,~IEEE}
\thanks{This work was partially supported by the Natural Sciences and Engineering Research Council of Canada. 
This work was partially supported by the National Natural Science Foundation of China under Grant 62203086 and 62173062. 
This work was partially supported by partially supported by the Fundamental Research Funds for the Central Universities under Grant 044420250104. {\itshape(Corresponding author: Qing Zhao).}}
\thanks{Z. Lv, Y. Liu, A. F. Lynch, and Q. Zhao are with Dept. of Elec. and Comp. Eng. Univ. of Alberta, Edmonton, Alberta, Canada. (e-mail: $<$zongyan3, yo12, alynch, qingz$>$@ualberta.ca).}
\thanks{Y. Jia is with School of Science, Dalian Minzu University, Dalian 116600, China (e-mail: jym@dlnu.edu.cn).}
\thanks{Y. H. Wu is with Key Laboratory of Intelligent Control and Optimization for Industrial Equipment of Ministry of Education, and School of Control Science and Engineering, Dalian University of Technology, Dalian, 116024, China. (e-mail: wuyuhu@dlut.edu.cn).}
}
\markboth{}
{Shell \MakeLowercase{\textit{et al.}}: Bare Demo of IEEEtran.cls for IEEE Journals}

\maketitle

\begin{abstract}
Unmanned aerial vehicle (UAV) with slung load system is a classic air transportation system. 
In practical applications, the suspension point of the slung load does not always align with the center of mass (CoM) of the UAV due to mission requirements or mechanical interference. 
This offset creates coupling in the system's nonlinear dynamics which leads to a complicated motion control problem. 
In existing research, modeling of the system are performed about the UAV's CoM. In this work we use the point of suspension instead. 
Based on the new model, a cascade control strategy is developed. 
In the middle-loop controller, the acceleration of the suspension point is used to regulate the swing angle of the slung load without the need for considering the coupling between the slung load and the UAV. 
An inner-loop controller is designed to track the UAV's attitude without the need of simplification on the coupling effects. 
We prove local exponential stability of the closed-loop using Lyapunov approach. 
Finally, simulations and experiments are conducted to validate the proposed control system. 
\end{abstract}


\IEEEpeerreviewmaketitle

\section{Introduction}
\IEEEPARstart{N}{owdays}, with the rapid developments of electronics and control science, unmanned aerial vehicles (UAVs) have seen widespread applications in various fields, such as aerial photography, infrastructure inspection, and wildfire monitoring \cite{mahony2012multirotor,10538180}. 
Since UAVs are not constrained by terrain, their application in transportation offers great convenience and versatility. 
In particular, UAV slung load systems have gained attention recently due to their benefits. 
These systems have a simple and reliable mechanical structure and can accommodate a wide range of loads, and unload safely without landing. 
As one of the important aerial transportation systems, the UAV with a slung load system has gained attention in the research field due to their distinctive advantages, including the ability to load and unload loads without landing, flexible volume constraints on the load, and structural simplicity of the suspension \cite{lv2022fixed}. 

Prior studies have made the critical assumption that the suspension point of the slung load coincides with the Center of Mass (CoM) of the UAV or have approximated the coupling effects as the external disturbances 
\cite{10068260,8910429,7843619,lv2021adaptive,cabecinhas2019trajectory,10106041,9104868,9749960}. 
However, due to operational requirements and mechanical constraints, the load's suspension point is rarely at the UAV's CoM. 
The misalignment of the suspension point from the UAV's CoM brings several significant challenges to the flight control design for a UAV with off-center slung load (UOSL). 
The suspended load effectively behaves like a double-pendulum system, and have a complex nonlinear dynamics. 
For a UAV with a slung load system without off-center property, the UAV's attitude dynamics is not influenced by the slung load. 
In contrast, for a UOSL system, the off-center slung load induces additional dynamic instabilities and introduces  nonlinear coupling between the UAV and slung load attitude dynamics. 

The research on control of UOSL is fairly limited. 
Zeng and Sreenath derived a dynamic model for the UOSL in a coordinate-free manner using Lagrange-d'Alembert principle and developed a geometric controller to track the UAV attitude, swing angle, and load position \cite{8814939}. 
The controller design assumes the angular acceleration of the UAV is negligible in order to simplify the dynamics.
However, for fast UAV motion and large offsets, this simplification does not hold and the coupling dynamics cannot be neglected for accurate motion control. 
Qian and Liu obtained the dynamic model of the UOSL using the  Kane's method and developed a controller to stabilize the UAV using partial linearization \cite{7946750}. The stability of control design in \cite{7946750} is not proven and based on a geometric control for a bare UAV \cite{5717652}. 
In all the aforementioned studies, both the dynamic model and control law were formulated with respect to a coordinate frame centered at the UAV's CoM. 
Furthermore, no existing strategy has fully addressed the coupling effects between the swing angle and UAV attitude without relying on simplification. 
Finally, to the best of our knowledge, no published studies have reported real-world flight experiments on the UOSL system. 


Motivated by the aforementioned issues and challenges, this work is focused on flight control design of UOSL system. 
Unlike classic centroid-based modeling approaches, we change the perspective from the UAV's CoM to the load tether point, from which we can derive a new dynamic model for the UOSL. 
Based on the new dynamic model, we propose a cascade control strategy: an outer-loop tension force control for the load liner velocity, a middle-loop acceleration control of the suspension point for the swing angle, an inner-loop off-center torque control for the UAV's attitude, and finally an off-center mixer. 
Compared with existing methods, the contributions of this work are summarized as follows:\\
1) In this study, we derive a new dynamic model for the UOSL 
using a reference frame at the suspension point, which offers a novel perspective on the dynamic coupling between the slung load and the UAV. 
This model reveals that the motion of the slung load is directly driven by the acceleration of the suspension point, and the UAV's attitude dynamics is not explicitly included in the dynamic model of the linear velocity and swing angle of the slung load.\\ 
2) Based on the constructed model, we design a nonlinear acceleration control law to actively control the motion of the load without the need to consider the coupling between the UAV and the slung load. 
The inner-loop attitude control torque, on the other hand, fully takes into account this coupling effect, and is designed to track the desired UAV attitude. 
Moreover, the inner-loop control law enables the estimation of the tension force on the slung load without the need for any additional force sensor. 
By the Lyapunov approach, the closed-loop system is proved to be locally exponentially stable theoretically.\\ 
3) The effectiveness of the proposed control strategy has been verified through real flight experiments. \\

The remainder of this paper is organized as follows. The dynamic model of the UOSL is established and the control problem is defined in Section II. 
The control strategy, as well as the stability analysis, is provided in Section III. 
In Section IV, simulations and experimental results are presented to demonstrate the performance of the proposed controller based on the new model of the UOSL. 
Finally, Section V concludes this paper and discusses the future work.

\section{Dynamic Modeling From Off-center Perspective}
For multirotor UAVs, such as quadrotor and hexacopters, aside from the mixer, their controller mostly share the same design principle. 
Without loss of generality, this paper is focused on the modeling and control design of the UOSL. 
In this work, we assume that the cable is inextensible. 
The symbols $\bm{s}$ and $\bm{c}$ are used to replace $\sin$ and $\cos$, respectively. 
The notation $\bm{0}_{m \times n}$ represents an $m \times n$  zero matrix, and $\bm{I}_n$ denotes an identity matrix.
%

\begin{figure}[ht]
\centering
\vspace{-0.3cm} 
\setlength{\belowcaptionskip}{-0.2cm}
\includegraphics[width=0.8\linewidth]{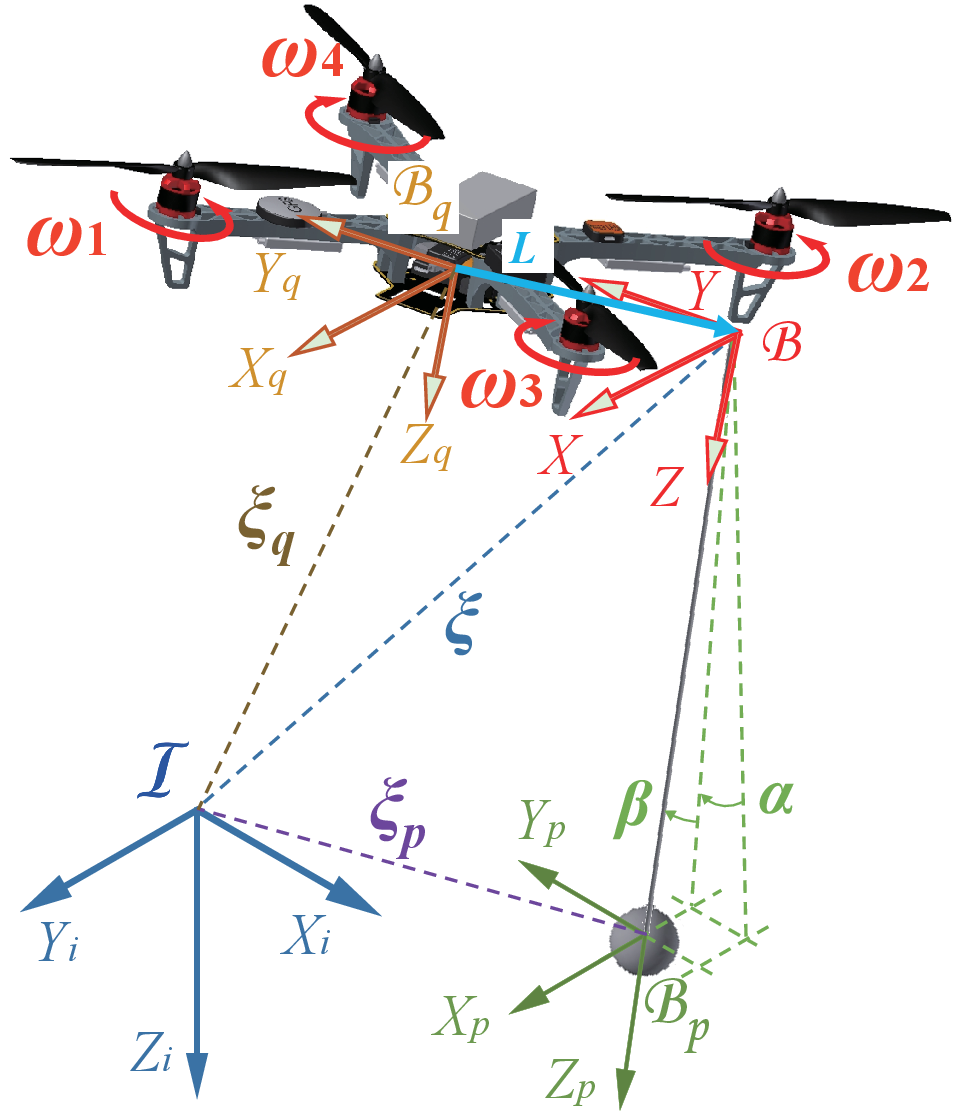}
\caption{\label{structure} The quadrotor UAV with an off-center slung load.}
\belowcaptionskip=-3pt
\end{figure}

The structure and coordinate frames of the UOSL are shown in Fig. \ref{structure}. 
The following coordinate frames are used to describe the UOSL: the inertial frame $\mathcal{I}\{\overrightarrow{X}_i, \overrightarrow{Y}_i, \overrightarrow{Z}_i\}$ following the North-East-Down (NED) notation; 
the load's body frame $\mathcal{B}_p\{\overrightarrow{X}_p, \overrightarrow{Y}_p, \overrightarrow{Z}_p\}$; 
the quadrotor's body frame $\mathcal{B}_q\{\overrightarrow{X}_q, \overrightarrow{Y}_q, \overrightarrow{Z}_q\}$. 
The axes of $\mathcal{B}_p$ and $\mathcal{B}_q$ are oriented forward, right, and down, respectively. 
The orientation of $\mathcal{B}$ is aligned with that of  $\mathcal{B}_q$, with its origin at the suspension point. 
Based on the aforementioned frames, the following state variables are defined: 
$\bm{\xi_q}=[x_q~y_q~z_q]^\top\in\mathbb{R}^3$, coordinate of the positions of the UAV's CoG in the frame $\bm{\mathcal{I}}$; 
$\bm{\eta}=[\phi~\theta~\psi]^\top\in \mathbb{R}^3$, the Euler angle of the quadrotor UAV; 
$\bm{\sigma}=[\alpha~\beta]^\top\in\mathbb{R}^2$, the swing Euler angle, where $\alpha$ and $\beta$ are the roll angle and pitch angle of the slung payload, respectively. 
Then, the generalized coordinate is defined as\\
$\bm{q}={[\bm{\xi_q}^\top~\bm{\eta}^\top~\bm{\sigma}^\top]^\top}\in\mathbb{R}^8$.
The present work exclude aggressive maneuvering, with the Euler angles bounded as follows: 
\begin{equation}\label{limitation}
\phi, \theta, \alpha, \beta \in(-\pi/2,\pi/2).
\end{equation}

The positions of the suspension point and the load's CoG are expressed by $\bm{\xi},\bm{\xi_p}\in\mathbb{R}^3$: 
\begin{subequations}\label{translation}
\begin{align}
	\bm{\xi} =& \bm{\xi_q}+\bm{R_{b}^{i}}\bm{L},\label{xiq}\\
	\bm{\xi_p} =& \bm{\xi_q}+\bm{R_{b}^{i}}\bm{L}+\bm{R_{p}^{i}}\bm{l}, \label{xip}
\end{align}
\end{subequations}
where $\bm{L}=[l_x~l_y~l_z]^\top$ is the offset vector from the origin of $\mathcal{B}_q$ to suspension point in the frame $\mathcal{B}_q$, 
$\bm{l}=[0~0~l]^\top$ is the vector from the tether point to the load's CoG in the frame $\mathcal{B}_p$, 
with the length of cable $l$ and the transition matrices\\
$
\bm{R_b^i}=(\bm{R_i^{b}})^{\top}=
\left[
\begin{array}{ccc}
\bm{c}\theta\bm{c}\psi& \bm{s}\phi\bm{s}\theta\bm{c}\psi-\bm{c}\phi\bm{s}\psi& \bm{c}\phi\bm{s}\theta\bm{c}\psi+\bm{s}\phi\bm{s}\psi\\
\bm{c}\theta\bm{s}\psi& \bm{s}\phi\bm{s}\theta\bm{s}\psi+\bm{c}\phi\bm{c}\psi& \bm{c}\phi\bm{s}\theta\bm{s}\psi-\bm{s}\phi\bm{c}\psi\\
-\bm{s}\theta& \bm{s}\phi\bm{c}\theta& \bm{c}\phi\bm{c}\theta
\end{array}\right], $
\\
$
\bm{R_{p}^{i}}=(\bm{R_i^{p}})^{\top}=\left[
\begin{array}{ccc}
\bm{c}\beta& \bm{s}\alpha\bm{s}\beta& \bm{c}\alpha\bm{s}\beta\\
0& \bm{c}\alpha& -\bm{s}\alpha\\
-\bm{s}\beta& \bm{s}\alpha\bm{c}\beta& \bm{c}\alpha\bm{c}\beta
\end{array}\right].$

The kinetic energy $T(\bm{q},\bm{\dot{q}})$ of the whole UOSL can be partitioned by 
$T(\bm{q},\bm{\dot{q}})=T_{qt}+T_{qr}+T_{pt}$, where $T_{qt}$ and $T_{qr}$ are the translational kinetic energy and the rotational kinetic energy of the UAV, respectively, $T_{pt}$ are the load's translational energy. 
The translational kinetic energies of the UAV and the load are given by 
\begin{subequations}\label{Tqt}
\begin{align}
T_{qt}=&\frac{1}{2}m_q\bm{\dot{\xi}_q}^{\top}\bm{\dot{\xi}_q},\\
T_{pt}=&\frac{1}{2}m_p\bm{\dot{\xi}_p}^{\top}\bm{\dot{\xi}_p},
\end{align}
\end{subequations}
where $m_q$ and $m_p$ denote the masses of the UAV and the load, respectively. 
The UAV's rotational kinetic energy is given by
\begin{align}\label{Tqr}
T_{qr}&=\frac{1}{2}\bm{\dot{\eta}}^\top\bm{J_q}\bm{\dot{\eta}},
\end{align}
where 
$\bm{J_q}=[J_{kj}]_{3\times3}=\bm{R_{v}}^{\top}\bm{I_q}\bm{R_{v}}$, $\bm{I_q}=\text{diag}(I_{qxx},I_{qyy},I_{qzz})$ denotes the UAV's rotational inertial matrix, 
with 
$$
\bm{R_{v}}=
\left[\begin{array}{ccc}
1&0&-\bm{s}\theta\\
0&\bm{c}\phi&\bm{s}\phi\bm{c}\theta\\
0&-\bm{s}\phi&\bm{c}\phi\bm{c}\theta
\end{array}\right].
$$
The potential energy of the UOSL is formulated as:
\begin{equation}\label{V}
V(\bm{q})=-m_q g z_q - m_p g z_p,
\end{equation}
where $g$ is the acceleration of gravity. 
Combining the energy components from (\ref{Tqt}), (\ref{Tqr}), and (\ref{V}), we establish the system's Lagrangian as 
\begin{equation} \label{Lagrangian}
L(\bm{q},\bm{\dot{q}})=T(\bm{q},\bm{\dot{q}})-V(\bm{q}).
\end{equation}

The dynamic model of the UOSL is established through the Euler-Lagrange formulation as follows:
\begin{equation} \label{EL}
\frac{d}{dt}\frac{\partial L(\bm{q},\bm{\dot{q}})}{\partial\bm{\dot{q}}}-\frac{\partial L(\bm{q},\bm{\dot{q}})}{\partial\bm{q}}=\bm{F_g},
\end{equation}
where $\bm{F_g}=\bm{F_a}+\bm{F_d}$ is the generalized external force with the  generalized control force $\bm{F_a}$ and the generalized drag force $\bm{F_d}$.  
Substituting the Lagrangian \eqref{Lagrangian} into \eqref{EL} yields 
\begin{equation} \label{M}
\bm{M}(\bm{q})\bm{\ddot{q}}+\bm{C}(\bm{q},\bm{\dot{q}})\bm{\dot{q}}+\bm{G}(\bm{q})=\bm{F_a}+\bm{F_d},
\end{equation}
where $\bm{M}(\bm{q})=\bm{M}(\bm{q})^\top=[m_{kj}]_{8\times 8}$ is a symmetric matrix.  

Following the method proposed in \cite{spong2020robot}, the elements of $\bm{C}(\bm{q},\bm{\dot{q}})=[\bm{C_\xi}^\top~\bm{C_\eta}^\top~\bm{C_\sigma}^\top]^\top=[c_{kj}]_{8\times 8}$ are calculated as:
$$
c_{kj}= \sum_{i=1}^{8}\Big(\frac{\partial m_{kj}}{\partial q_i}+\frac{\partial m_{ki}}{\partial q_j}-\frac{\partial m_{ij}}{\partial q_k} \Big)\frac{\dot{q}_i}{2},
$$
with  
$\bm{C_\xi},\bm{C_\eta}\in\mathbb{R}^{3\times8}$, and
$\bm{C_\sigma}\in\mathbb{R}^{2\times8}$. 
The elements of vector $\bm{G}(\bm{q})=[\bm{G_\xi}^\top~\bm{G_\eta}^\top~\bm{G_\sigma}^\top]^\top$ are obtained as follows:
\begin{align*}
	\bm{G_\xi}&\!=[0~0~-(m_q + m_p)g]^\top,\\
	\bm{G_\eta}&\!=m_pg[-\bm{c}\theta(\bm{c}\phi l_y \!-\!\bm{s}\phi l_z)~~
	\!\big(\bm{c}\theta l_x \!+\! \bm{s}\theta(\bm{s}\phi l_y \!+ \!\bm{c}\phi l_z)\big)~0]^\top,\\
	\bm{G_\sigma}&\!=m_pg[\bm{s}\alpha\bm{c}\beta l ~\bm{c}\alpha\bm{s}\beta l]^\top.
\end{align*}

The generalized control force in (\ref{M}) is formulated as
\begin{equation*}\label{Fa}
\bm{F_a}=\bm{B}\bm{u}, 
\end{equation*}
where $\bm{u}=[F_l~\bm{\tau_\eta}^\top]^\top$ represents the control input of the UOSL system, including the thrust force $F_l$ and control torque $\bm{\tau_\eta}$ generated by the rotor. 
The control effectiveness matrix $\bm{B}$ is given by
$\bm{B}=\left[\begin{array}{ccc}
\bm{R}&\bm{0}_{3\times3}\\
\bm{0}_{3\times1}&\bm{I}_{3}\\
\bm{0}_{2\times1}&\bm{0}_{2\times3}\\
\end{array}\right],$
with $\bm{R}=\bm{R_b^i}[0~0~1]^\top$.

The generalized drag force in (\ref{M}) is expressed as
\begin{equation*}\label{Fd}
\bm{F_d}=[\bm{D_{\xi q}}^\top+\bm{D_{\xi p}}^\top~~\bm{D_\eta}^\top~~ \bm{D_\sigma}^\top]^\top,
\end{equation*}
where $\bm{D_{\xi q}}$ and $\bm{D_{\xi p}}$ represent the aerodynamic drag forces acting on the quadrotor and load, respectively.
The term $\bm{D_\eta}$ denotes the torque produced by the air resistance.
The values of $\bm{D_{\xi q}}$, $\bm{D_{\xi p}}$ and $\bm{D_\eta}$ can be calculated by the method proposed in \cite{lv2020nonlinear}.
The torque $\bm{D_\sigma}=[D_\alpha~D_\beta]^\top$, generated by air resistance on the slung load, is given by 
$$\bm{D_\sigma}=
\begin{bmatrix}
	1 & 0 & 0 \\
	0 & 1 & 0
\end{bmatrix}
\bigg(\bm{l}\times\bm{R_i^{p}}\bm{D_{\xi p}}\bigg).$$

Then, the dynamic model of the UOSL in \eqref{M} is proposed as the control-oriented form as follows:
\begin{subequations}\label{model}
	\begin{align}
		&\bm{\ddot{\xi}_p}=(\bm{F_t} +m_p\bm{g} +\bm{D_{\xi p}})/m_p,\label{model_xi1}\\
		&\bm{\ddot{\eta}}=\bm{J_{q}}^{-1}(\bm{\tau_\eta}+\bm{\tau_{Ft}}-\bm{\tilde{C}_\eta}\bm{\dot{\eta}}+\bm{D_{\eta}}), \label{model_eta1}\\
		&\bm{\ddot{\sigma}} =-\bm{M_{\sigma}}\bm{\ddot{\xi}}-\bm{M_{\sigma 1}}^{-1} (\bm{\tilde{C}_\sigma}\bm{\dot{\tilde{q}}}+\bm{G_\sigma}-\bm{D_{\sigma}}),\label{model_sigma1}
	\end{align}
\end{subequations}
In \eqref{model_xi1}, 
$\bm{F_t}=\!\bm{R} \bm{F_l}-m_q\bm{\ddot{\xi}_q}+m_q\bm{g}+\bm{D_{\xi q}}$ is the tensile force on the cable, and $\bm{g}=\![0~0~g]^\top$.  
In \eqref{model_eta1},
\begin{equation}\label{tauq}
	\bm{\tau_{Ft}}=\bm{R_v}^{\top}\big(-\bm{L}\times\bm{R_b^i}\bm{F_t} \big), 
\end{equation}
is the torque generated by the tensile force. 
The term $\bm{\tilde{C}_\eta}=[\tilde{c}_{kj}]_{3\times3}$ is  the Coriolis torque of the UAV, with 
$$\tilde{c}_{kj}=\sum_{i=1}^{3}\Big(\frac{\partial J_{kj}}{\partial \eta_i}+
\frac{\partial J_{ki}}{\partial \eta_j} -\frac{\partial J_{ij}}{\partial \eta_k}\Big)\frac{\dot{\eta_i}}{2}.$$
Then, the model \eqref{model_sigma1} is obtained by substituting the relationship \eqref{xiq} into the last two equations of the model \eqref{M},  
with \eqref{model_sigma1},
$\bm{\tilde{C}_\sigma}=\bm{C_\sigma}\, \text{diag}(\bm{I}_3,\bm{0}_{3\times 3},\bm{I}_2)$,
$\bm{\tilde{q}}={[\bm{\xi}^\top~\bm{\eta}^\top~\bm{\sigma}^\top]^\top}$, $\bm{M_{\sigma}}=\bm{M_{\sigma 1}}^{-1}\bm{M_{\sigma 2}}$,
$\bm{M_{\sigma1}}=
\begin{bmatrix}
	m_{77} & 0 \\
	0 & m_{88}
\end{bmatrix}
$,
$\bm{M_{\sigma2}}=
\begin{bmatrix}
	m_{71} & m_{72} & m_{73} \\
	m_{81} & 0 & m_{83}
\end{bmatrix}
$.

\begin{remark}
Differing from the existing model of the UOSL \cite{8814939,7946750}, 
(\ref{model}) is constructed based on an off-center perspective. 
This model reveals that the load's swing angle is driven by the acceleration $\bm{\ddot{\xi}}$ of the suspension point (\ref{model_sigma1}). 
Furthermore, the load's swing dynamics (\ref{model_sigma1}) does not have the terms coupled with the UAV attitude $\bm{\eta}$ explicitly. 
\end{remark}

\section{Control Design}
It should be noted that the UOSL has four control inputs (four rotors) but eight degrees of freedom. Hence, it is an underactuated system. We propose a cascade control structure in this paper. The goal is to track the load velocity. 
\begin{figure*} 
\centering
\vspace{0.1cm} 
\setlength{\belowcaptionskip}{-0.4cm}
\includegraphics[width=0.8\textwidth]{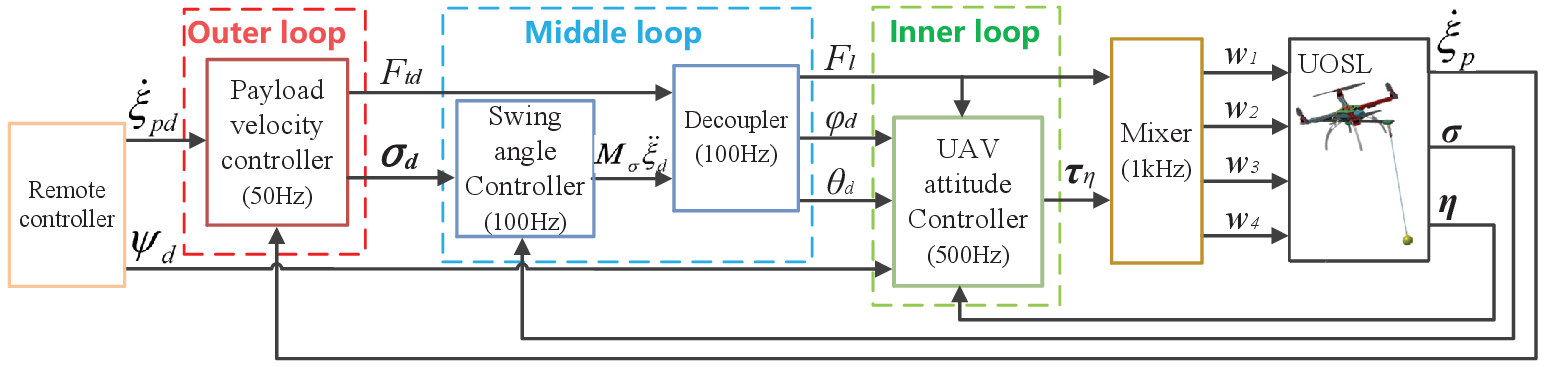}
\caption{The schematic of the control strategy.}
\label{fig:controlSchematic}
\end{figure*}

\subsection{Mixer}
The control input $\bm{u}=[F_l~~\bm{\tau_\eta}^\top]^\top$ in (\ref{M}) 
is realized by the thrust and reaction torque produced by the rotors.
According to \cite{martini2024correction}, $(\bm{R_{v}}^{\top})^ {-1}\bm{\tau_\eta}=\bm{\tau_b}=[\tau_x~\tau_y~\tau_z]^\top$. 
Then, the forces and torque on the quadrotor can be written as
\begin{equation*}
\left[\!\!
\begin{array}{c}
	F_l\\
	{\tau}_{x} \\
	{\tau}_{y} \\
	{\tau}_{z}
\end{array}
\!\!\right]\!=\!
\left[\!\!
\begin{array}{cccc}
	-1 & -1 & -1 & -1 \\
	-l_{ra}\!\! & l_{ra}\!\! & l_{ra}\!\! & -l_{ra}\\
	l_{ra}\!\! & -l_{ra}\!\! & l_{ra}\!\! & -l_{ra}\\
	c_q & c_q & -c_q & -c_q
\end{array}
\!\right]
\left[\!
\begin{array}{c}
	F_1 \\
	F_2 \\
	F_3 \\
	F_4
\end{array}
\!\right],
\end{equation*}
where $l_{ra}=l_r/\sqrt{2}$, with the distance $l_r$ from the rotational axes of the rotors to the CoM of the quadrotor, $F_i~(i=1_{\cdots} 4)$ is the thrust generated by rotor $i$, and $c_q$ denotes the torque coefficient of rotors. 
Then, the mixer of the quadrotor can be solved as follows: 
\begin{equation*}
\left[\!\!
\begin{array}{c}
	F_1 \\
	F_2 \\
	F_3 \\
	F_4
\end{array}
\!\!\right]=
\left[\!\!
\begin{array}{l}
	((-\!\sqrt{2}/2F_l-\tau_x+ \tau_y)c_q/2+\!\sqrt{2}\tau_z/4)/l_{rb}\\
	((-\!\sqrt{2}/2F_l+\tau_x-\tau_y)c_q/2+\!\sqrt{2}\tau_z/4)/l_{rb}\\
	((-\!\sqrt{2}/2F_l+\tau_x+ 
	\tau_y)c_q/2-\!\sqrt{2}\tau_z/4)/l_{rb}\\
	((-\!\sqrt{2}/2F_l-\tau_x- 
	\tau_y)c_q/2-\!\sqrt{2}\tau_z/4)/l_{rb}
\end{array}
\!\!\right],
\end{equation*}
where $l_{rb}=\sqrt{2}c_q l_{r}$. 


\subsection{Cascade control system}
The proposed cascade control structure consists of the inner-loop attitude controller, the middle-loop swing angle controller, and the outer-loop load velocity controller, which are discussed in the following. 

Before proposing the control scheme, we define tracking error variables.
The configuration errors for different variables are defined. 
Given desired load velocity $\bm{\dot{\xi}_{pd}}=[\dot{x}_{pd}~\dot{y}_{pd}~\dot{z}_{pd}]^\top$, quadrotor attitude $\bm{\eta_d}=[\phi_d~\theta_d~\psi_d]^\top$, and swing angle $\bm{\sigma_d}=[\alpha_d~\beta_d]^\top$, the error system is defined as follows: 
\begin{subequations}\label{error}
\begin{align}
	&\bm{e_{\dot{\xi}p}}=\bm{\dot{\xi}_{pd}}-\bm{\dot{\xi}_p}, \label{exip}\\
	&\bm{e_\eta}=\bm{\eta_d}-\bm{\eta},~
	\bm{e_{p_\eta}}=\bm{\dot{\eta}_d}-\bm{\dot{\eta}}+\bm{K_\eta}\bm{e_\eta},~\bm{e_{\eta,{p_\eta}}}=[\bm{e_\eta}^\top~\bm{e_{p_\eta}}^\top]^\top,\label{eeta}\\
	&\bm{e_\sigma}=\!\bm{\sigma_d}\!-\!\bm{\sigma},~
	\bm{e}_{\bm{p_\sigma}}=\bm{\dot{\sigma}_d}-\!\bm{\dot{\sigma}}+\bm{K_\sigma}\bm{e_\sigma},~\!\bm{e_{\sigma,{p_\sigma}}}=[\bm{e_\sigma}^\top~\bm{e_{p_\sigma}}^\top]^\top,
 \label{esigma}
\end{align}
\end{subequations}
with the positive definite diagonal matrixes $\bm{K_\eta}=\text{diag}(k_{\phi},k_{\theta},k_{\psi})$ and $\bm{K_\sigma}=\text{diag}(k_{\alpha},k_{\beta})$. 
Taking the time derivative of $\bm{e_\eta}$ and $\bm{e_\sigma}$ yields
\begin{subequations}\label{ed}
\begin{align}
	&\bm{\dot{e}}_{\bm{\eta}}=\bm{e}_{\bm{p_\eta}}-\bm{K_\eta}\bm{e_\eta},~
	\bm{\dot{e}_{p_{\bm{\eta}}}}=\bm{\ddot{\eta}}_d-\bm{\ddot{\eta}}+\bm{K_\eta}(\bm{e_{p_\eta}}-\bm{K_\eta} \bm{e_\eta}),\label{edeta}\\
	&\bm{\dot{e}}_{\bm{\sigma}}=\bm{e}_{\bm{p_\sigma}}-\bm{K_\sigma}\bm{e_\sigma},~
	\bm{\dot{e}_{p_{\bm{\sigma}}}}=\bm{\ddot{\sigma}}_d-\bm{\ddot{\sigma}}+\bm{K_\sigma}(\bm{e_{p_\sigma}}-\bm{K_\sigma}\bm{e_\sigma}).\label{edsigma}
\end{align}
\end{subequations}

\subsubsection{Inner-loop Attitude Controller}
In designing the inner-loop attitude controller and solving the decoupling dynamics between the slung load and the UAV, a critical step is the computation of the suspension point's acceleration $\bm{\ddot{\xi}}$, which is simultaneously influenced by both the thrust $F_l$ and the UAV attitude $\bm{\eta}$. 
At first, we design the expected UAV attitude convergent  rate in advance as 
\begin{equation}\label{acc_eta}
\bm{\ddot{\eta}_{tr}}=(\bm{I}_3-\bm{K_{\eta}}^2) \bm{e_{\eta}}+(\bm{K_{\eta}}+\bm{K_{p{\eta}}}) \bm{e_{p_\eta}}, 
\end{equation}
where $\bm{K_{p\eta}}\!=\!\text{diag}(k_{p\phi},k_{p\theta},k_{p\psi})$ is a constant positive definite matrix.
According to (\ref{translation}) and (\ref{model_xi1}), the relationship equation between the UAV and the suspended load is established through the tension $F_t$ on the cable as follows
\begin{align}\label{Ft}
\bm{F_t}=
&\bm{R}F_l-m_q\big(\bm{\ddot{\xi}}-\bm{\ddot{R}_{b}^{i}}(\bm{\ddot{\eta}_{tr}})\bm{L}\big)+m_q\bm{g}+\bm{D_{\xi q}}\notag\\ =&\bm{R_{p}^i}[0~0~F_t]^\top=m_p\bm{\ddot{\xi_p}} -m_p\bm{g}-\bm{D_{\xi p}}, 
\end{align}
where $\bm{\xi_p} = \bm{\xi}+\bm{R_{p}^{i}}\bm{l}$.  
Based on \eqref{Ft}, we have
\begin{align}\label{xi_calculation}
\bm{\ddot{\xi_p}}\!=&
\Big(\bm{R}F_l \!+\!m_q\big(\bm{\ddot{R}_{b}^{i}}(\bm{\ddot{\eta}_{tr}})\bm{L} \!+\!\bm{\ddot{R}_{p}^{i}}\bm{l}\big) \!+\!\bm{D_{\xi q} \!+\!\bm{D_{\xi p}}}\Big)/(m_q\!+\!m_p)+\bm{g}.
\end{align}
Substituting the thrust force $F_l$  generated by the middle-loop swing angle controller in \eqref{solveFld1-Fl} and the prescribed trajectory (\ref{acc_eta}), the acceleration of the load $\bm{\ddot{\xi_p}}$ can be computed using \eqref{xi_calculation}. Subsequently, the tensile force vector $\bm{F_t}$ and its magnitude $F_t$ can be obtained by substituting $\bm{\ddot{\xi_p}}$ into \eqref{Ft},
and the control torque $\bm{\tau_{Ft}}$ in \eqref{tauq} can be obtained. 

Building on the aforementioned results, the inner-loop attitude controller is designed as follows:
\begin{align}\label{taueta}
\bm{\tau_\eta}\!=\!\bm{J_{q}}(\bm{\ddot{\eta}_{tr}}+\bm{\ddot{\eta}_{d}})-\bm{\tau_{Ft}}+\bm{\tilde{C}_\eta}\bm{\dot{\eta}}-\bm{D_{\eta}}. 
\end{align}

\begin{theorem}
For the quadrotor attitude dynamics in (\ref{model_eta1}), if the torque $\bm{\tau_\eta}$ is set as (\ref{taueta}), the zero equilibrium of the attitude tracking errors $\bm{e_\eta}$ and $\bm{e_{p_\eta}}$ are locally exponentially stable. 
\end{theorem}
\begin{IEEEproof}
The Lyapunov candidate $V_{\bm{\eta}}$ is constructed as
\begin{equation}\label{Veta}
V_{\bm{\eta}}(\bm{e_{\eta,{p_\eta}}})=\|\bm{e_{\eta,{p_\eta}}}\|^2/2.
\end{equation}
Substituting (\ref{edeta}) into time derivative of Lyapunov candidate $V_{\bm{\eta}}$ in (\ref{Veta}) results in
\begin{align*}
\dot{V}_{\bm{\eta}}=&\bm{e_\eta}^\top(\bm{e_{p_\eta}}-\bm{K_\eta} \bm{e_\eta})+\bm{e_{p_\eta}}^\top[\bm{\ddot{\eta}}_d-\bm{\ddot{\eta}}+\bm{K_\eta}(\bm{e_{p_\eta}}-\bm{K_\eta} \bm{e_\eta})]\\
=&\bm{e_{\eta,{p_\eta}}}^\top \bm{\dot{e}_{\eta,{p_\eta}}}.
\end{align*}
Implementing the attitude controller (\ref{taueta}) in the dynamic model (\ref{model_eta1}) yields
$
\bm{\ddot{\eta}}=\bm{\ddot{\eta}_{tr}}+\bm{\ddot{\eta}_{d}}. 
$
Then, we have
\begin{align*}\label{dVeta:2}
\dot{V}_{\bm{\eta}}=&\bm{e_\eta}^\top(\bm{e_{p_\eta}}\!-\!\bm{K_\eta} \bm{e_\eta})
	\!+\!\bm{e_{p_\eta}}^\top[(\bm{K_\eta}^2\!-\!\bm{I}_3)\bm{e_\eta}\!-\!(\bm{K_\eta}\!+\!\bm{K_{p\eta}})\bm{e_{p_\eta}}\nonumber\\
	&+\bm{K_\eta}(\bm{e_{p_\eta}}-\bm{K_\eta}\bm{e_\eta})]\nonumber\\
	=&-\bm{e_{\eta,{p_\eta}}}^\top\bm{W_\eta}\bm{e_{\eta,{p_\eta}}}
	\leq-\lambda_{min}(\bm{W_\eta})\|\bm{e_{\eta,{p_\eta}}}\|^2,
\end{align*}
where
$\bm{W_\eta}=\text{diag}(\bm{K_\eta},\bm{K_{p\eta}})$ is positive definite because of the positive definite property of matrixes $\bm{K_\eta}$ and $\bm{K_{p\eta}}$. 
$\lambda_{min}(\cdot)$ denotes the minimum eigenvalue of a matrix.

Recalling (\ref{Veta}) and letting $\lambda_{\bm{\eta}}=2\lambda_{min}(\bm{W_\eta}),$ we have
\begin{equation*}\label{Veta:inequality}
	\dot{V}_{\bm{\eta}}\leq-\lambda_{\bm{\eta}}V_{\bm{\eta}}.
\end{equation*}
Consequently, the zero equilibrium of the attitude tracking errors $\bm{e_\eta}$ and $\bm{e_{p_\eta}}$ of the inner-closed-loop system (\ref{model_eta1}) and (\ref{taueta}) are locally exponentially stable.
\end{IEEEproof}

\subsubsection{Middle-loop Swing Angle Controller}
The swing angle controller is applied to track desired swing angle $\bm{\sigma}_d$. 
The term $\bm{M_{\sigma}}\bm{\ddot{\xi}}$ with $\bm{M_{\sigma}}=\bm{M_{\sigma 1}}^{-1}\bm{M_{\sigma 2}}$ is taken as the virtual  control input of the middle-loop system, and the desired control law is designed as 
\begin{align}\label{Fsigmad}
\bm{M_{\sigma}}\bm{\ddot{\xi}_d} =& -\Big(\bm{\ddot{\sigma}_d}+(\bm{I}_2-\bm{k_\sigma}^2)\bm{e_\sigma}+(\bm{k_\sigma}+\bm{k_{p\sigma}})\bm{e_{p\sigma}}\Big)\nonumber\\
&-\bm{M_{\sigma 1}}^{-1}(\bm{C_\sigma}\bm{\dot{q}}+\bm{G_\sigma}-\bm{D_{\sigma}}),
\end{align}
where $\bm{k_{p\sigma}}=\text{diag}(k_{p\alpha},k_{p\beta})$ is a constant positive definite matrix. 

\begin{theorem}
Given the desired swing angle $\bm{\sigma}_d$, 
if the virtual control input $\bm{M_{\sigma 2}}\bm{\ddot{\xi}}$
is chosen as (\ref{Fsigmad}), 
the zero equilibria of the swing angle tracking errors $\bm{e_\sigma}$ and $\bm{e_{p_\sigma}}$  of the system (\ref{model_sigma1}) are locally exponentially stable. 
\end{theorem}
\begin{IEEEproof}
The Lyapunov candidate $V_{\sigma}$ is designed as 
\begin{equation}\label{Vsigma}
	V_{\sigma}=\|\bm{e_{\sigma,{p_\sigma}}}\|^2/2.
\end{equation}

Define the control input error as 
\begin{align}\label{emsigmaddxi}
	\bm{e_{M\sigma \xi}}=\bm{M_{\sigma}}\bm{\ddot{\xi}}_d-\bm{M_{\sigma}}\bm{\ddot{\xi}},
\end{align}
where
$\bm{M_{\sigma}}\bm{\ddot{\xi}}_d$ is given in (\ref{Fsigmad}).
Substituting the control input error $\bm{e_{M\sigma \xi}}$ in (\ref{emsigmaddxi}) into (\ref{model_sigma1}) yields 
\begin{align}\label{ddsigma}
	\bm{\ddot{\sigma}_d}-\bm{\ddot{\sigma}}=&(\bm{k_\sigma}^2-\bm{I}_2)\bm{e_\sigma}-(\bm{k_\sigma}+\bm{k_{p\sigma}})\bm{e_{p_\sigma}}-\bm{e_{M\sigma \xi}}.
\end{align}
According to (\ref{edsigma}),
the time derivative of $\bm{e_{\sigma,{p_\sigma}}}$ in (\ref{esigma}) is obtained as
\begin{equation}\label{desigmapsigma}
	\bm{\dot{e}_{\sigma,{p_\sigma}}}(\bm{e_{\sigma,{p_\sigma}}},\bm{e_{M\sigma \xi}})=\left[ \begin{array}{c}
		\bm{e_{p_\sigma}}-\bm{k_\sigma}\bm{e_\sigma}\\
		\bm{\ddot{\sigma}}_d-\bm{\ddot{\sigma}}+\bm{k_\sigma}(\bm{e_{p_\sigma}}-\bm{k_\sigma}\bm{e_\sigma})
	\end{array}
	\right].
\end{equation}

Then, substituting (\ref{ddsigma}) and (\ref{desigmapsigma}) into the time derivative of Lyapunov candidate $V_{\sigma}$ in (\ref{Vsigma}) yields
\begin{align}\label{dVsigma}
	\dot{V}_{\sigma}=&(\partial V_{\sigma}/\partial \bm{e_{\sigma,{p_\sigma}}})\bm{\dot{e}_{\sigma,{p_\sigma}}}(\bm{e_{\sigma,{p_\sigma}}},\bm{e_{M\sigma \xi}})\nonumber\\
	=&\bm{e_\sigma}^\top(\bm{e_{p_\sigma}}-\bm{k_\sigma}\bm{e_\sigma})+\bm{e_{p_\sigma}}^\top\Big((\bm{k_\sigma}^2-\bm{I}_2)\bm{e_\sigma}-(\bm{k_\sigma}+\bm{k_{p\sigma}})\bm{e_{p_\sigma}}\nonumber\\
	&-\!\bm{e_{M\sigma \xi}}+\bm{k_\sigma}(\bm{e_{p_\sigma}}\!-\!\bm{k_\sigma}\bm{e_\sigma})\Big).
\end{align}
When the actual control law $\bm{M_{\sigma}}\bm{\ddot{\xi}}$ is set as $\bm{M_{\sigma}}\bm{\ddot{\xi}}_d$ in (\ref{Fsigmad}), which means $\bm{e_{M\sigma \xi}}=0$, we have 
\begin{align}\label{dVsigma:1}
	\dot{V}_{\sigma}=&\bm{e_\sigma}^\top(\bm{e_{p_\sigma}}-\bm{k_\sigma}\bm{e_\sigma})+\bm{e_{p_\sigma}}^\top[(\bm{k_\sigma}^2-\bm{I}_2)\bm{e_\sigma}-(\bm{k_\sigma}+\bm{k_{p\sigma}})\bm{e_{p_\sigma}}\nonumber\\
	&+\bm{k_\sigma}(\bm{e_{p_\sigma}}-\!\bm{k_\sigma}\bm{e_\sigma})]\nonumber\\
	=&-\bm{e_{\sigma,{p_\sigma}}}^\top\bm{W_\sigma}\bm{e_{\sigma,{p_\sigma}}}\leq-\lambda_\sigma V_{\sigma}\leq0.
\end{align}
where
$\bm{W}_{\bm{\sigma}}\!=\!\text{diag}(\bm{k_\sigma},~\bm{k_{p\sigma}})$
and $\lambda_\sigma\!=\!2\lambda_{min}(\bm{W_\sigma})$ are positive definite. 

Consequently, the zero equilibria of the swing errors $\bm{e_\sigma}$ and $\bm{e_{p_\sigma}}$ of the middle-closed-loop system (\ref{model_sigma1}) and (\ref{Fsigmad}) are locally exponentially stable.
\end{IEEEproof}

\subsubsection{Decoupler}
For the desired virtual control input $\bm{M_\sigma}\bm{\ddot{\xi}}_d\in\mathbb{R}^2$ in (\ref{Fsigmad}) generated by the aforementioned swing angle controller and the desired tension force $F_{td}$ obtained from the outer-loop velocity controller, 
the decoupler is utilized to calculate the thrust $F_l$ and the desired attitude $\phi_d$, $\theta_d$ by decoupling $\bm{M_\sigma}\bm{\ddot{\xi}}_d$ and $F_{td}$. 
The procedure is presented as follows. 

Since the cable is inelastic and in steady-state,  $\bm{\ddot{\xi}_q}$ is equal to the acceleration $\bm{\ddot{\xi}}$ of the tether point along axis $Z_b$ at steady state, we obtain the following equation: 
\begin{equation}\label{Solveddxi:1}
	[0~0~1]\bm{R_i^{p}}\bm{\ddot{\xi}}_d=\kappa,
\end{equation}
where 
$$
\kappa=\Big(F_{td}+[0~0~1]\bm{R_i^{p}}(\bm{D_{\xi p}}+m_p\bm{g})\Big)/m_p.
$$
Define the virtual control input as 
\begin{equation}\label{ddsigma_v}
	\bm{M_{\sigma}}\bm{\ddot{\xi}_d} =-\bm{\ddot{\sigma}_{v}}=-[\ddot{\alpha}_{v}~\ddot{\beta}_{v}]^\top.  
\end{equation}
Combing equations (\ref{Solveddxi:1}) and (\ref{ddsigma_v}), the acceleration $\ddot{\bm{\xi}}_d=[\ddot{x}_d~\ddot{y}_d~\ddot{z}_d]^\top$  can be solved as 
\begin{subequations} \label{Solveddxi}
	\begin{align}
		\ddot{x}_d=&\bm{c}\alpha\bm{s}\beta\kappa - l\bm{c}\alpha\bm{c}\beta \ddot{\beta}_v+l\bm{s}\alpha\bm{s}\beta\ddot{\alpha}_v,\\
		\ddot{y}_d=&l\bm{c}\alpha\ddot{\alpha}_v-\bm{s}\alpha\kappa,\\
		\ddot{z}_d=&\bm{c}\alpha\bm{c}\beta\kappa + l\bm{c}\alpha\bm{s}\beta \ddot{\beta}_v+l\bm{s}\alpha\bm{c}\beta\ddot{\alpha}_v.
	\end{align}
\end{subequations}
Based on $\ddot{\bm{\xi}}_d$ in (\ref{Solveddxi}) and the (\ref{model_xi1}), the desired thrust $\bm{F_{ld}}=[F_{lxd}~F_{lyd}~F_{lzd}]^\top$ can be calculated as
\begin{equation}\label{Fld}
	\bm{F_{ld}}=m_q\bm{\ddot{\xi}}_d
	+\bm{R}_i^{bp\top}\left[0~0~F_{td}\right]^\top
	-m_q \bm{g}-\bm{D_\xi}.
\end{equation}
Considering the thrust limitation of the rotors, we design the constraint on the desired thrust 
$\bm{F_{ld}}$ as follows: 
\begin{align*}
\begin{split}
    \bm{F_{ld}^r}=\left \{
    \begin{array}{ll}
        \![0~0~-F_{up}]^\top, & \!\!\text{if}~F_{lzd} <-F_{up}, \\  
        \!{[hF_{lxd}~hF_{lyd}~F_{lzd}]}^\top, & \!\!\text{if}~F_{lzd}\geq -F_{up}~\text{\&}~||\bm{F_{ld}}||>F_{up}, \\	
        \!\bm{F_{ld}},  & \!\!\text{if}~||\bm{F_{ld}}||\leq F_{up},
    \end{array}
    \right.
\end{split}
\end{align*}
where $\bm{F_{ld}^r}=[F_{lxd}^r~F_{lyd}^r~F_{lzd}^r]^\top$ denotes the desired constrained thrust, with $F_{lzd}^r$ constrained by $F_{lzd}^r<0$, $F_{up}$ is the upper bound of the thrust,  $h=\sqrt{F_{up}^2-F_{lzd}^2}/\sqrt{||\bm{F_{ld}}||^2-F_{lzd}^2}$.  
It should be noted that the upper bound $F_{up}$ is state-dependent, and its real-time calculation is computationally involved. 
In this work, it is assumed as a constant value based on the performance of the rotors. 
This assumption is satisfied for most practical UOSL motions.

Given the relationship in (\ref{Fa}), the desired lift force $\bm{F_{ld}^r}$ is decoupled into the total thrust $F_l$ generated by the rotors and the desired swing angle $\phi_d$, $\theta_d$ using:
\begin{equation}\label{solveFld}
	{
		\left[\!\begin{array}{ccc}
			\bm{c}\psi\!&\!-\bm{s}\psi\!&\!0\\
			\bm{s}\psi\!&\!\bm{c}\psi\!&\!0\\
			0\!&\!0\!&\!1\\
		\end{array}\!\right]\!\left[\!\begin{array}{ccc}
			\bm{c}\theta_d\!&\!0\!&\!\bm{s}\theta_d\\
			0\!&\!1\!&\!0\\
			-\bm{s}\theta_d\!&\!0\!&\!\bm{c}\theta_d
		\end{array}\!\right]\!
		\left[\!\begin{array}{ccc}
			1\!&\!0\!&\!0\\
			0\!&\!\bm{c}\phi_d\!&\!-\bm{s}\phi_d\\
			0\!&\!\bm{s}\phi_d\!&\!\bm{c}\phi_d
		\end{array}\!\right]\!
		\left[\!\begin{array}{c}
			0\\0\\F_l
		\end{array}\!\right]
		=\bm{F_{ld}^r}.}
\end{equation}
Solving (\ref{solveFld}) yields 
\begin{subequations}\label{solveFld1}
\begin{align}
\theta_d=&\arctan\left((F_{lxd}^r\bm{c}\psi+F_{lyd}^r\bm{s}\psi)/F_{lzd}^r\right),\label{solveFld1-theta}\\
\phi_d=&-\arctan\left((-F_{lxd}^r\bm{s}\psi+F_{lyd}^r\bm{c}\psi)\bm{c}\theta_d/F_{lzd}^r\right),\label{solveFld1-phi}\\
F_{l}=&F_{lzd}^r/(\bm{c}\phi\bm{c}\theta).\label{solveFld1-Fl}
\end{align}
\end{subequations}

\subsubsection{Outer-loop Load Velocity Controller}
The desired outer-loop linear velocity controller $\bm{F_{td}}=[F_{txd}~F_{tyd}~F_{tzd}]^\top$ is designed based on the dynamic model (\ref{model_xi1}) as
\begin{equation}\label{Ftd}
\bm{F_{td}}=\bm{k_{\dot{\xi}p}}\bm{e_{\dot{\xi}p}}+m_p \bm{\ddot{\xi}_{pd}}+\bm{C_\xi}\bm{\dot{q}}-m_p\bm{g}-\bm{D_{\xi p}}, 
\end{equation}
with a positive definite matrix $\bm{k_{\dot{\xi}p}}=\text{diag}(k_{\dot{x}_p},k_{\dot{y}_p},k_{\dot{z}_p})$.

\begin{theorem}
Given a desired load velocity $\bm{\dot{\xi}_{pd}}$, if the tension $\bm{F_t}$ is chosen as $\bm{F_{td}}$ in (\ref{Ftd}),  the zero equilibria of the velocity tracking error $\bm{e_{\dot{\xi}p}}$ of the closed-loop system (\ref{model_xi1}) and (\ref{Ftd}) is locally exponentially stable. 
\end{theorem}
\begin{IEEEproof}
	Define control input error as 
	\begin{equation}\label{eFt}
		\bm{e}_{\bm{F_t}}=\bm{F_{td}}-\bm{F_t}.
	\end{equation}
	Substituting the dyanmic model (\ref{model_xi1}) into  (\ref{exip}) yields 
	\begin{equation}\label{esigma:equation}
		\bm{\dot{e}_{\dot{\xi}p}}=\bm{\ddot{\xi}_{pd}}-(\bm{F}_t-\bm{C_\xi}\bm{\dot{q}} +m_p\bm{g}+\bm{D_{\xi p}})/m_p.
	\end{equation}
	Then, substituting (\ref{Ftd}) and (\ref{eFt}) into  (\ref{esigma:equation}) yields
	\begin{equation*}
		\bm{\dot{e}_{\dot{\xi}p}}=-(\bm{k_{\dot{\xi}p}}\bm{e_{\dot{\xi}p}}-\bm{e}_{\bm{F_t}})/m_p.
	\end{equation*}
Setting $\bm{F}_{t}$ as $\bm{F_{td}}$ in (\ref{Ftd}), which means $\bm{e_{Ft}}=0$, then, we have
\begin{equation}\label{esgima:equation2}
	\bm{\dot{e}_{\dot{\xi}p}}=-\bm{k_{\dot{\xi}p}}\bm{e_{\dot{\xi}p}}/m_p.
\end{equation}
Consequently, the zero equilibrium of the velocity error $\bm{\dot{e}_{\dot{\xi}p}}$ of the outer-closed-loop system (\ref{model_xi1}) and (\ref{Ftd}) are locally exponentially stable.
\end{IEEEproof}

The relationship between the desired tension force vector $\bm{F_{td}}$, the magnitude of tension force $F_{td}$, and the swing angle $\bm{\sigma}$ is given as follows
\begin{equation}\label{solveFtd}
	\bm{F_{td}}=
	\bm{R_{pd}^i}
	\left[0~0~F_{td}\right]^\top,
\end{equation}
where
$
\fontsize{8}{8} \selectfont
\bm{R_{pd}^i}=\left[\!\begin{array}{ccc}
	\bm{c}\beta_d&0&\bm{s}\beta_d\\
	0&1&0\\
	-\bm{s}\beta_d&0&\bm{c}\beta_d
\end{array}\!\right]
\left[\!\begin{array}{ccc}
	1&0&0\\
	0&\bm{c}\alpha_d&-\bm{s}\alpha_d\\
	0&\bm{s}\alpha_d&\bm{c}\alpha_d
\end{array}\!\right].
$
Then, the three unknown variables, $F_{td}$, $\alpha_d$, and $\beta_d$ in (\ref{solveFtd}) can be solved as follows
\begin{subequations}\label{Ftd:solve}
\begin{align}
F_{td}&=F_{tzd}/(\bm{c}\alpha_d \bm{c}\beta_d),\label{Ftd:solve_value}\\
\beta_d&=\arctan(F_{txd}/F_{tzd}),\label{Ftd:solve_beta}\\
\alpha_d&=-\arctan(F_{tyd}\bm{c}\beta_d/F_{tzd}).\label{Ftd:solve alpha}
\end{align}
\end{subequations}
In this work, we do not consider scenarios where the slung load is invovled in aggressive vertical maneuvers, implying that $F_{tzd}<0$. Furthermore, given the constraints $\alpha_d, \beta_d \in (-\pi/2, \pi/2)$ specified in (\ref{limitation}), the solutions in (\ref{Ftd:solve}) are justifiable. 

Finally, for the entire closed-loop UOSL system, it can be shown that it is locally exponentially stable using the approach in \cite{lv2020nonlinear,Khalil2015dynamics}.

In summary, the proposed control scheme for the UOSL system includes the following steps:\\ 
1. The outer-loop velocity control law $\bm{F_{td}}$ given in (\ref{Ftd}) is used to track the desired load velocity $\bm{\dot{\xi}_{pd}}$;\\
2. The outer-loop control input $\bm{F_{td}}$ is transformed into the desired tension force  $F_{td}$ and the swing angle $\bm{\sigma_d}$ using (\ref{Ftd:solve});\\
3. The desired swing angle $\bm{\sigma_d}$  is tracked via the middle-loop virtual  control input $\bm{M_{\sigma}}\bm{\ddot{\xi}_d}$ in (\ref{Fsigmad});\\
4. The middle-loop control input   $\bm{M_{\sigma}}\bm{\ddot{\xi}_d}$ is transformed into the desired UAV attitude  $\phi_d$, $\theta_d$ and thrust $F_l$ using the decoupler in (\ref{Solveddxi}) and (\ref{solveFld1});\\
5.  The inner-loop controller provides $\bm{\tau_{\eta}}$ (\ref{taueta}) to track the desired attitude $\bm{\eta_d}$.


\begin{remark}
In this study, a model-based cascaded control framework is developed from an off-centered perspective. 
The middle-loop controller 
$\bm{M_{\sigma}}\bm{\ddot{\xi}_d}$ in (\ref{Fsigmad}) is designed to drive the swing angle and load linear velocity dynamics from an off-centered perspective, which does not explicitly include terms that coupled with the UAV attitude. 
All coupling terms are incorporated into the UAV's inner-loop attitude control law $\bm{\tau_\eta}$ in (\ref{taueta}), which includes the feedforward $\bm{M_{\eta 2}}\bm{\ddot{\xi}}$ and $\bm{G_{\eta}}$ to compensate for the torques induced by the UAV's inertia force and gravity. 
This independent design simplifies the control design and differs fundamentally from existing approaches, such as those in \cite{raffo2016nonlinear,sreenath2013geometric,9291453,9260232,Palunko2012a,8814939,9104868}. 
Typically, the method in \cite{8814939} assumes that the term $\hat{q}_u^2R_b^i\dot{\hat{\Omega}}L/l$ is negligible under the assumption of low UAV angular acceleration, where $\bm{q}_u$ denotes the unit vector from the suspension point to the load in the inertial frame $\mathcal{I}$, and $\bm{\Omega}$ represents the UAV's angular velocity. 
As demonstrated in Section IV, this approximation may have negative impact on the control performance.  
\end{remark}

\section{Simulation and Experimental Results}
\begin{figure}[ht]
	\centering
	\vspace{-0.3cm} 
	\setlength{\belowcaptionskip}{-0.2cm}
	\includegraphics[width=0.8\linewidth]{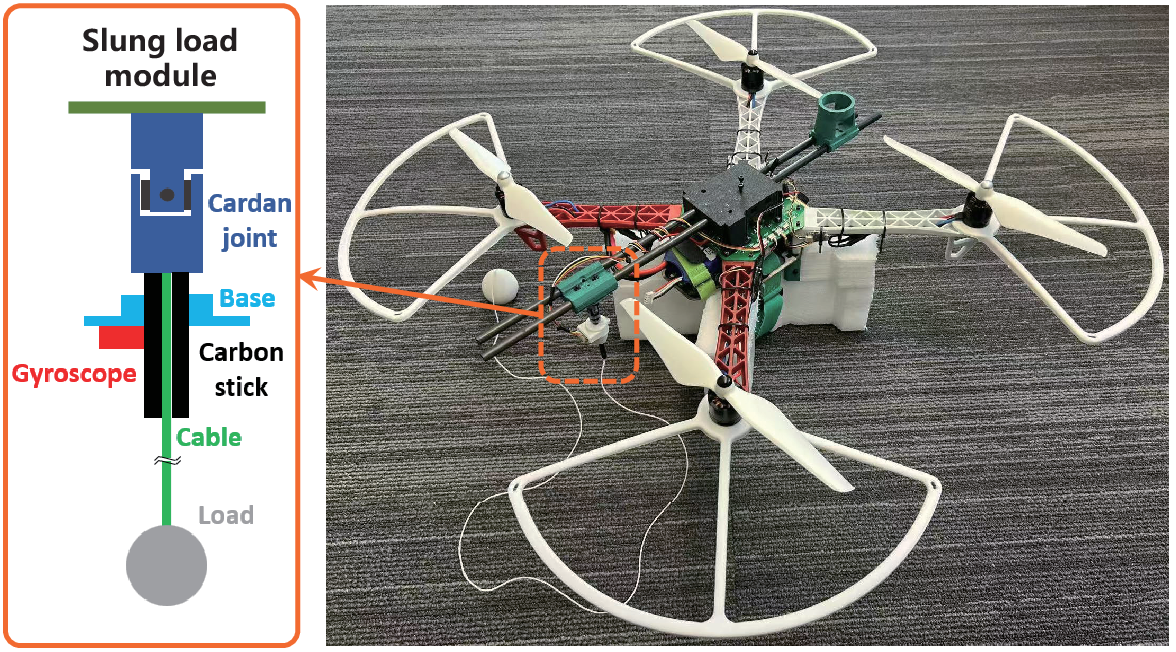}
	\caption{\label{platform}  Experimental platform.}
	\belowcaptionskip=-3pt
\end{figure}

To verify the effectiveness of the dynamic model and the designed control strategy, both simulations and experiments are performed. 
An experimental UOSL platform is developed based on that used  in \cite{lv2022finite}. 
The experimental platform is shown in Fig. \ref{platform}, and its physical parameters are listed in Table \ref{tab:parameter}. 
In real flight experiments, disturbances caused by rotor downwash acting on the off-center slung load introduce additional forces and torques. This can lead to small, high-frequency oscillations on the cable. 
Adding a streamlined shell around the load can help mitigate these effects. 
The terms $\bm{e_{p_\eta}}$ and $\bm{e_{p_\sigma}}$ in (\ref{acc_eta}) and (\ref{Fsigmad}) explicitly include the generalized velocity errors $\bm{\dot{\eta}_d}-\bm{\dot{\eta}}$, $\bm{\dot{\sigma}_d}-\bm{\dot{\sigma}}$ as well as the generalized position errors $\bm{e_{\eta}}$ and $\bm{e_{\sigma}}$. 
Given the PD terms of controller, the parameters of the control laws (\ref{acc_eta}), (\ref{Fsigmad}), and (\ref{Ftd}) are tuned using the Ziegler–Nichols method \cite{aastrom2004revisiting}.  
The control parameters of the simulations and experiments are 
$\bm{k_\eta}=\text{diag}(13.6,13.6,5.2)$, $\bm{k_{p\eta}}=\text{diag}(13.6,13.6,5.2)$, $\bm{k_\sigma}=\text{diag}(3.2,3.2)$, $\bm{k_{p\sigma}}=\text{diag}(3.2,3.2)$, and $\bm{k_{\dot{\xi}p}}=\text{diag}(1.4,1.4,4)$. 

In the manual mode, the control inputs generated by the remote controller cannot be obtained in advance. 
Therefore, the desired accelerations $\bm{\ddot{\eta}_d}$, $\bm{\ddot{\sigma}_d}$, and $\bm{\ddot{\xi}_{pd}}$ in the control laws (\ref{taueta}), (\ref{Fsigmad}), and (\ref{Ftd}) are set to zero in this work. 
Furthermore, velocity control is typically a fundamental control objective in manual operation and forms a critical foundation for higher-level trajectory tracking control. 
Thus, the experimental validation of this work is particularly focused on assessing the performance of the load linear velocity tracking. 

\begin{table}[H]
	\scriptsize
	\setlength{\belowcaptionskip}{-0.2cm}
	\centering
	\caption{Physical Parameters}
	\begin{tabular}{cccc}
		\hline\hline
		Parameter & Description & Value & Unit\\
		\hline
		$g$ & Acceleration of gravity & 9.81 & $\text{m/s}^2$\\
		$m_q$ & Mass of quadrotor & 1.32 & $\text{kg}$\\
		$m_p$ & mass of load & 0.066 & $\text{kg}$\\
		$l_r$ & Length of Rotor's arm & 0.225 & $\text{m}$\\
		$l$ & Cable length & 1 & $\text{m}$\\
		$I_{qxx}$, $I_{qyy}$  & Moment of inertia & $12.71\times10^{-3}$ & $\text{kg}\cdot \text{m}^2$ \\
		$I_{qzz}$& Moment of inertia & $2.37\times10^{-3}$& $\text{kg}\cdot \text{m}^2$ \\
		\hline\hline
	\end{tabular}
	\label{tab:parameter}
\end{table}

\subsection{Simulation}
\begin{figure}[!htb]
	\centering
	\vspace{-0.1cm} 
	\setlength{\belowcaptionskip}{-0.1cm}
	\includegraphics[width=0.95\linewidth]{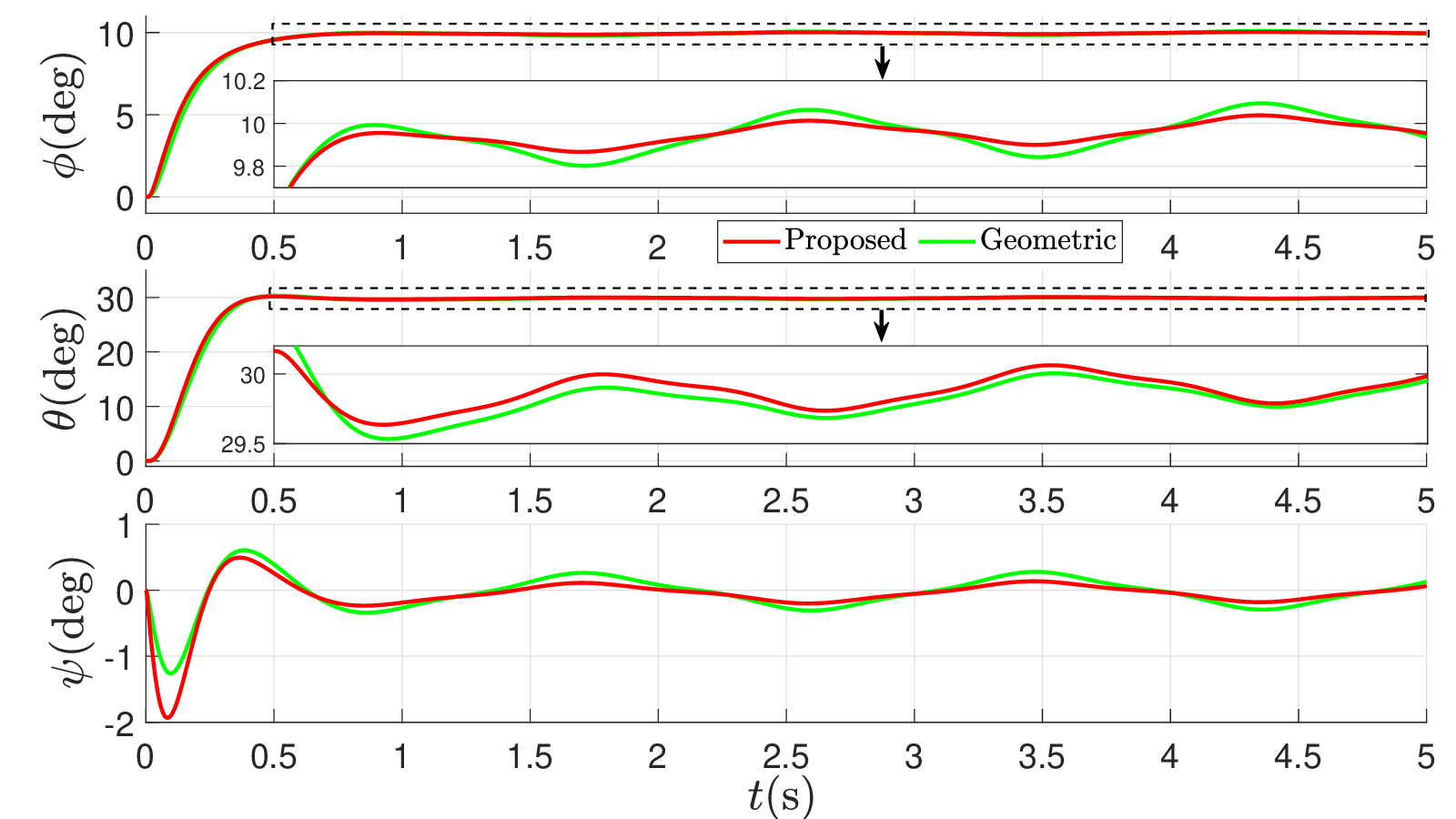}
	\caption{\label{figure:simulation}  Simulation results of UAV attitude tracking.}
	\belowcaptionskip=-3pt
\end{figure}

In this section, the comparison of the proposed scheme with the controller in \cite{8814939} is conducted. The model in \cite{8814939} neglects the coupling dynamics associated with the UAV's attitude acceleration, which may degrade the control performance. 
The measurement noise and unknown disturbances are inevitable in real flight and can affect the experimental results. 
Therefore, to clearly demonstrate the superiority of the proposed control strategy compared with the one in \cite{8814939}, we use a MATLAB/SimMechanics simulation environment. 
This environment provides an ideal and fair simulation platform, in which the UOSL model is generated based on a CAD design rather than simplified analytical formulations \cite{russell2018kinematics}. 
In the simulation tests, the UOSL tracks the desired UAV attitude $\bm{\eta} = [10~30~0]^\top~(\text{deg})$ from the initial state $\bm{\eta} = [0~0~0]^\top~(\text{deg})$ under different controllers, while the outer-loop and middle-loop controllers are deactivated. The corresponding results are shown in Fig.~\ref{figure:simulation}. It can be observed that the proposed controller achieves faster convergence and smaller tracking errors. 
The Root-Mean-Square Errors (RMSEs) of the Euler angles $\phi$, $\theta$, and $\psi$ obtained by the proposed controller are only $0.1395^\circ$, $0.0579^\circ$, and $0.0921^\circ$, respectively, which are $25.4\%$, $31.2\%$, $40.7\%$ lower than those of the controller in \cite{8814939} ($0.1869^\circ$, $0.0842^\circ$, and $0.1554^\circ$, respectively).


\subsection{Ground experiment}
\begin{figure}[!htb]%
\centering  %
\vspace{-0.2cm} %
\subfigtopskip=1pt %
\subfigbottomskip=1pt %
\setlength{\belowcaptionskip}{-0.2cm}  %
\subfigcapskip=-1pt %
\subfigure[UAV attitude]
{\label{figure:experiment_ground_rol}\includegraphics[width=0.95\linewidth]{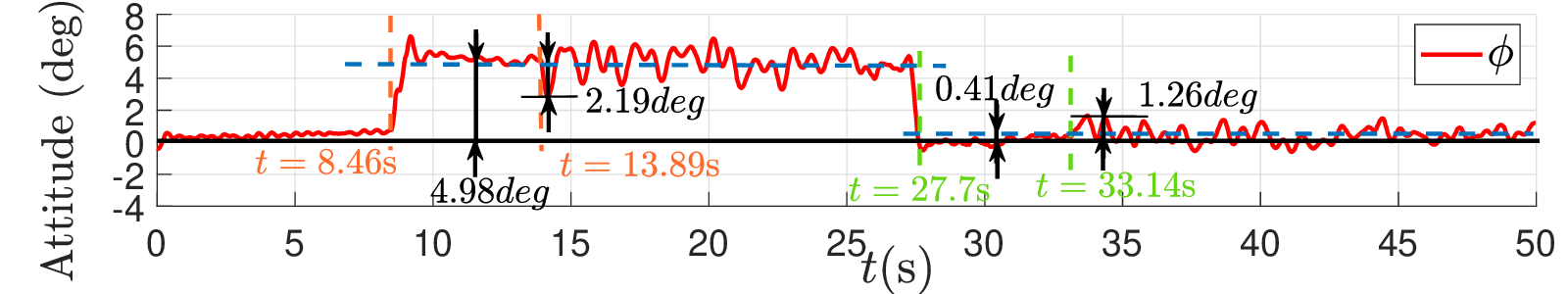}}
\subfigure[Control signals of rotors]
{\label{figure:experiment_ground_pwm}\includegraphics[width=0.95\linewidth]{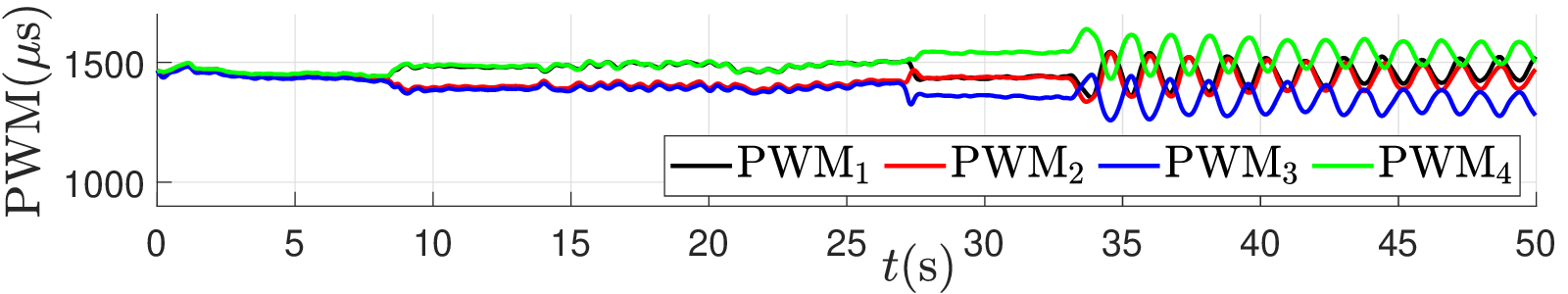}}
\subfigure[Ground test platform]
{\label{figure:ground_testbed}\includegraphics[width=0.8\linewidth]{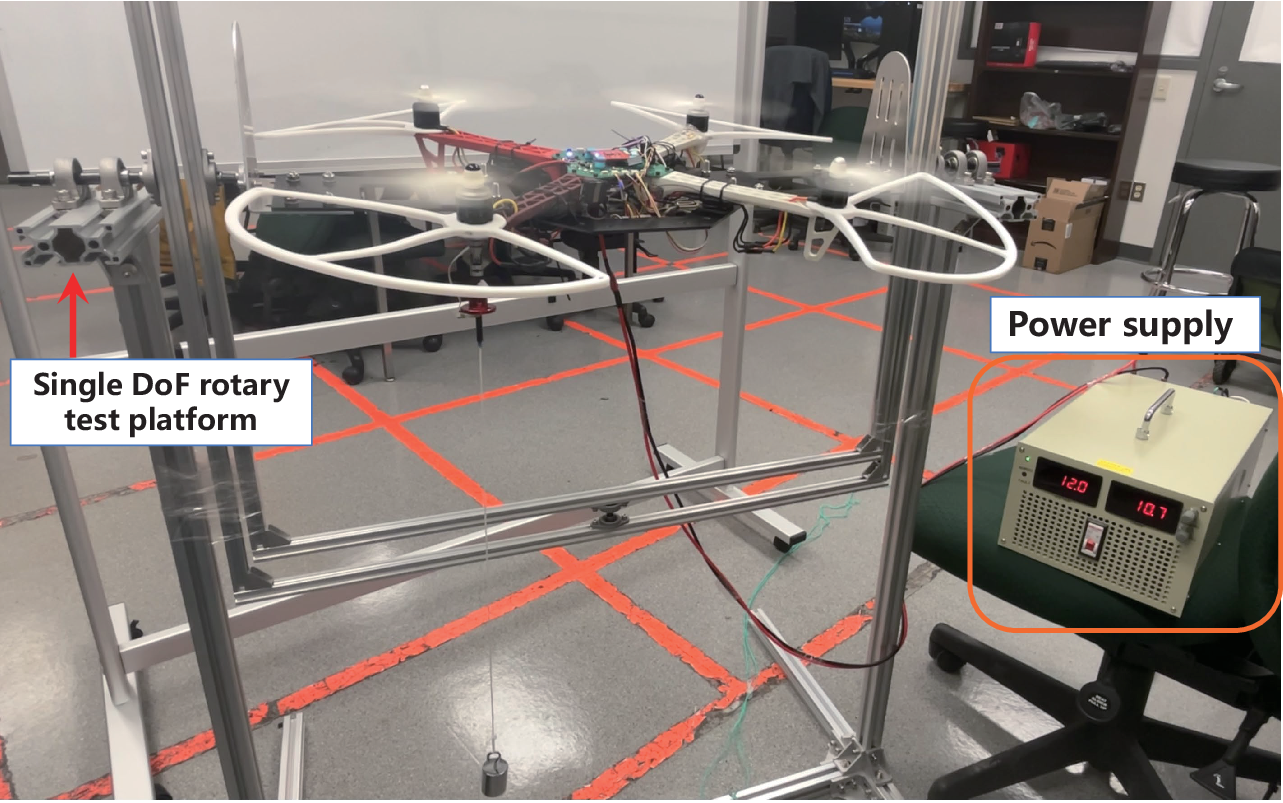}}
\caption{Ground test experiment.}
\label{figure:experiment_ground}
\end{figure}

In the ground test, the proposed controller is compared with the backstepping controller (BS) without considering the offset property. 
Considering the symmetry of the UAV, we only conduct roll motion experiments with offset $\bm{L}=[-0.159~0.159~0]^\top\text{m}$ to verify the effectiveness of the attitude controller. 
The UOSL is installed on a ground test bench that allows only for roll motion and try to track desired attitude $\phi_d=0^\circ$. 
The ground test platform is shown in Fig. \ref{figure:ground_testbed}

The results of the ground experiment are presented in Fig. \ref{figure:experiment_ground}. 
At first, the UOSL is stabilized by the BS controller. 
Between $t=5\text{s}$ and $t=11.66\text{s}$, a $0.05\text{kg}$ load is suspended at the cable without swing motion. 
The mean tracking error is $4.58^\circ$. 
At $t=11.66\text{s}$, a swinging motion is applied to the load, generating a varying disturbance torque. 
The maximum attitude oscillation with respect to its mean value reached $1.81^\circ$, and the standard deviation from $t=11.66\text{s}$ to $t=28.12\text{s}$ is $0.7529^\circ$. 
From $t=28.12\text{s}$, the load swing is manually suppressed, and the control of UOSL switches to the designed controller. 
We find that the attitude deviation is reduced by the proposed control strategy. 
At $t=36.53\text{s}$, a swinging motion is applied to the load. 
With the developed control method, the maximum attitude oscillation and the standard deviation are reduced to $1.51^\circ$ and $0.5414^\circ$, indicating the improvement of $16.57\%$ and $28.09\%$, receptivity, compared to the BS controller. 
These results confirm that the proposed control strategy can actively compensate for the influence caused by the suspended load and exhibits better robust performance than the BS controller. 
The video of the ground experiment is accompanied: 
\url{https://youtu.be/4hbEvUsaWFA}.

\subsection{Real flight experiment}
\begin{figure}[!htb]%
\centering  %
\vspace{-0.2cm} %
\subfigtopskip=1pt %
\subfigbottomskip=1pt %
\setlength{\belowcaptionskip}{-0.2cm}  %
\subfigcapskip=-1pt %
\subfigure[Load velocity and altitude]
{\label{figure:experiment_flight_velocity}\includegraphics[width=0.95\linewidth]{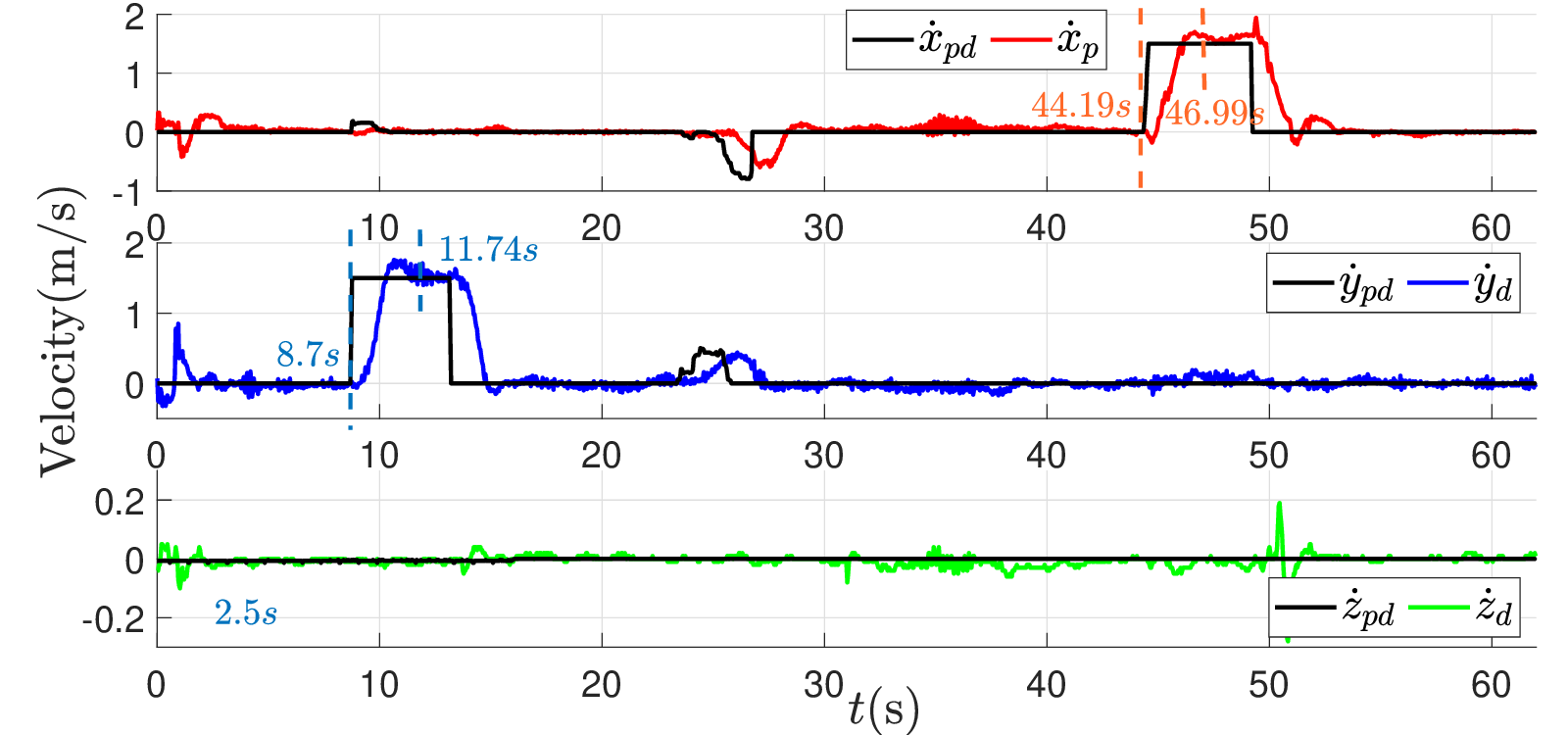}}
\subfigure[Swing angle]
{\label{figure:experiment_flight_swing}\includegraphics[width=0.95\linewidth]{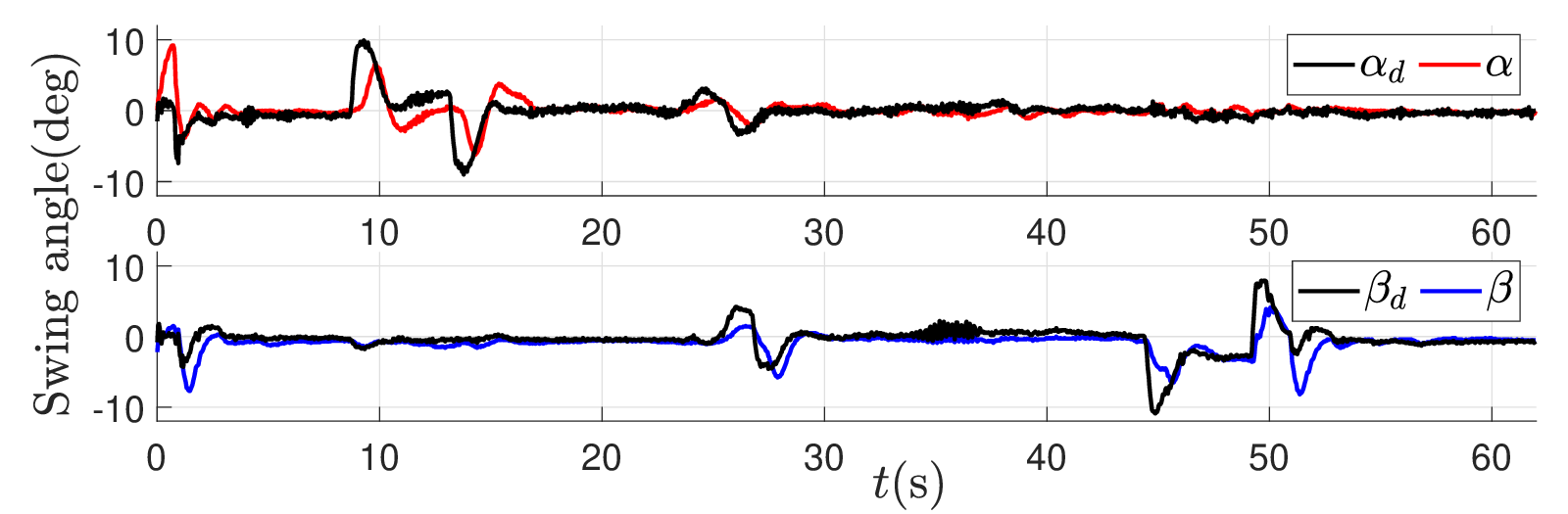}}
\subfigure[UAV attitude]
{\label{figure:experiment_flight_attitude}\includegraphics[width=0.95\linewidth]{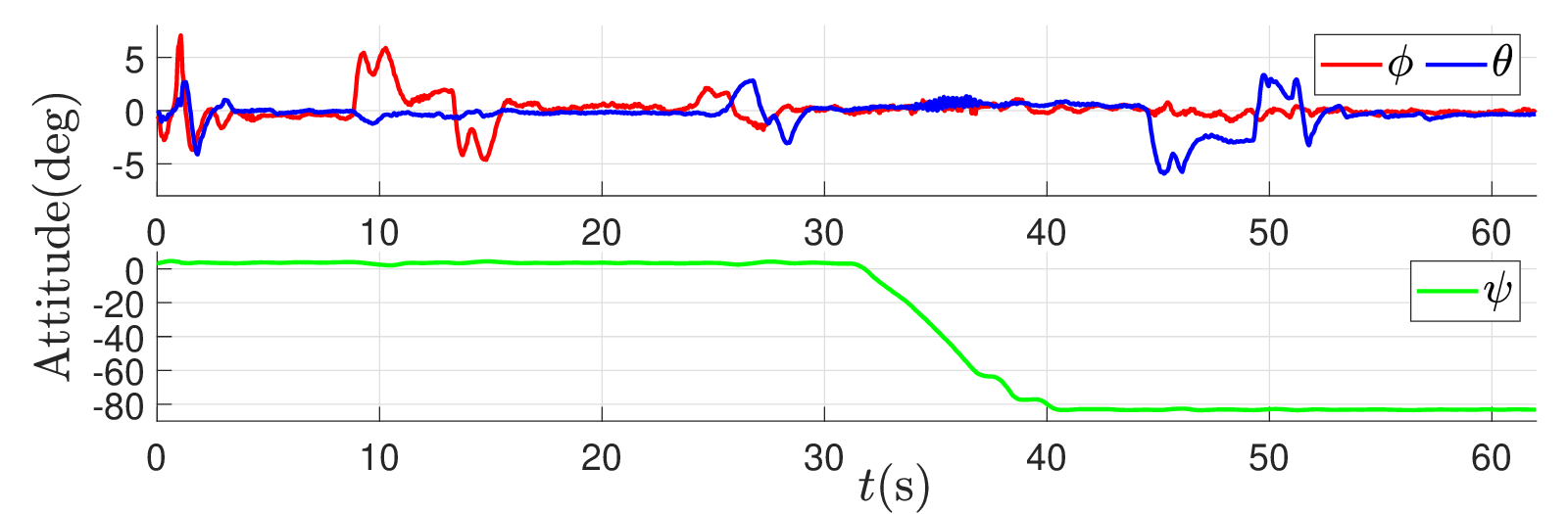}}
\subfigure[tension force estimation]
{\label{figure:experiment_flight_Ft}\includegraphics[width=0.95\linewidth]{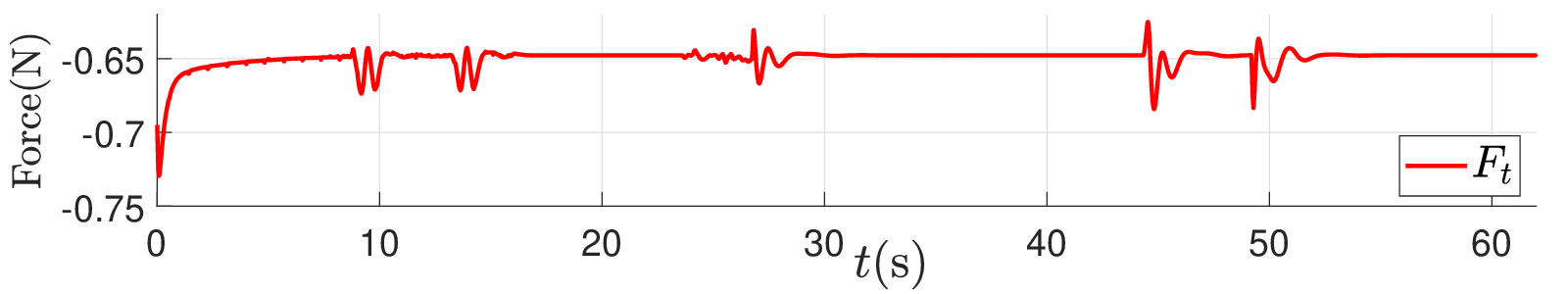}}
\subfigure[Control signals of rotors]
{\label{figure:experiment_flight_pwm}\includegraphics[width=0.95\linewidth]{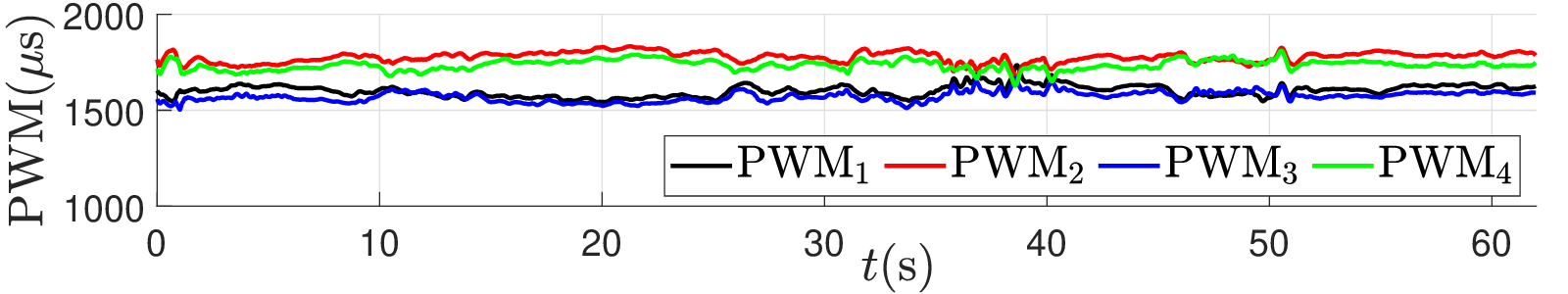}}
\caption{Results of flight experiment ($\bm{L}=[-0.12~0~-0.05]^\top m$).}
\label{figure:experiment_flight_12}
\end{figure}

\begin{figure}[!htb]%
\centering  %
\vspace{-0.2cm} %
\subfigtopskip=1pt %
\subfigbottomskip=1pt %
\setlength{\belowcaptionskip}{-0.2cm}  %
\subfigcapskip=-1pt %
\subfigure[Load velocity and altitude]
{\label{figure:experiment_flight_velocity_18}\includegraphics[width=0.95\linewidth]{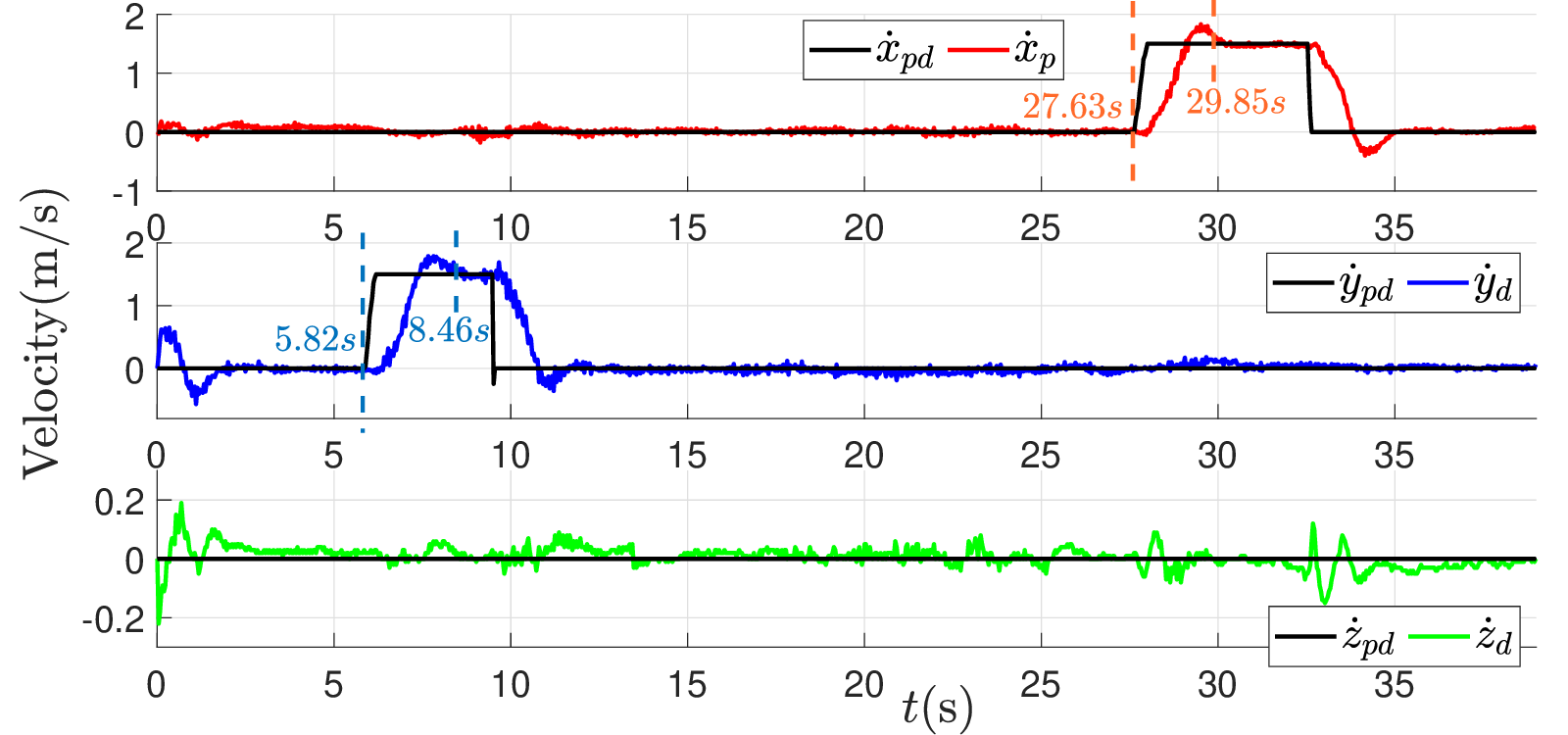}}
\subfigure[Swing angle]
{\label{figure:experiment_flight_swing_18}\includegraphics[width=0.95\linewidth]{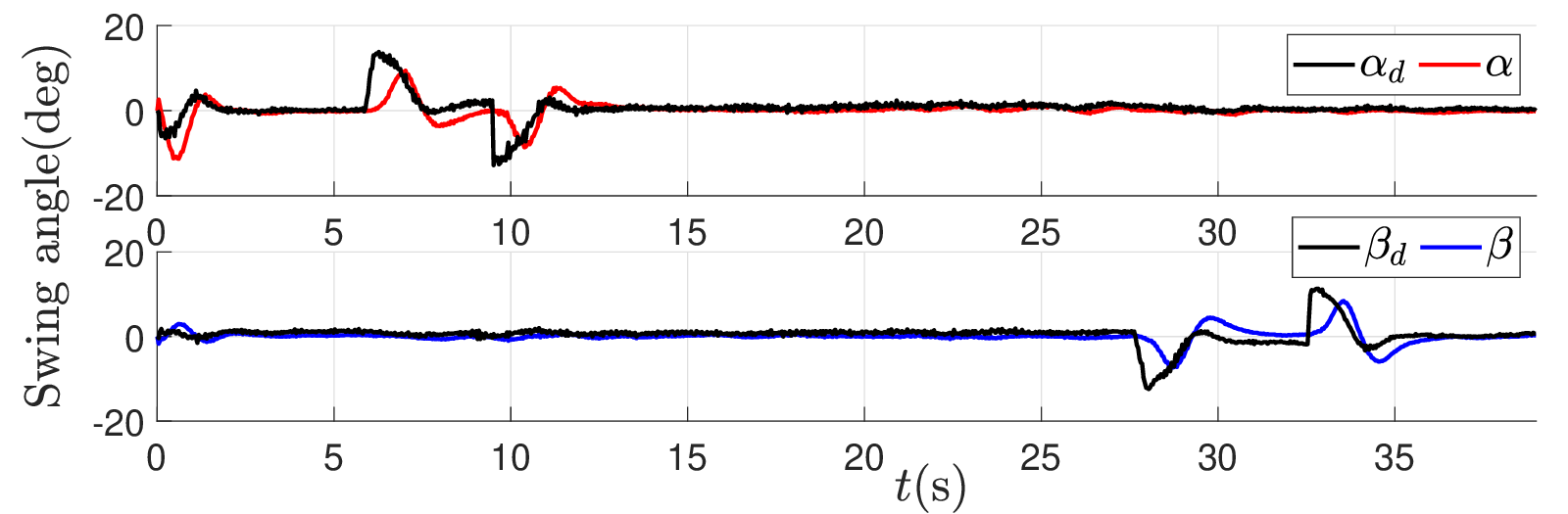}}
\subfigure[UAV attitude]
{\label{figure:experiment_flight_attitude_18}\includegraphics[width=0.95\linewidth]{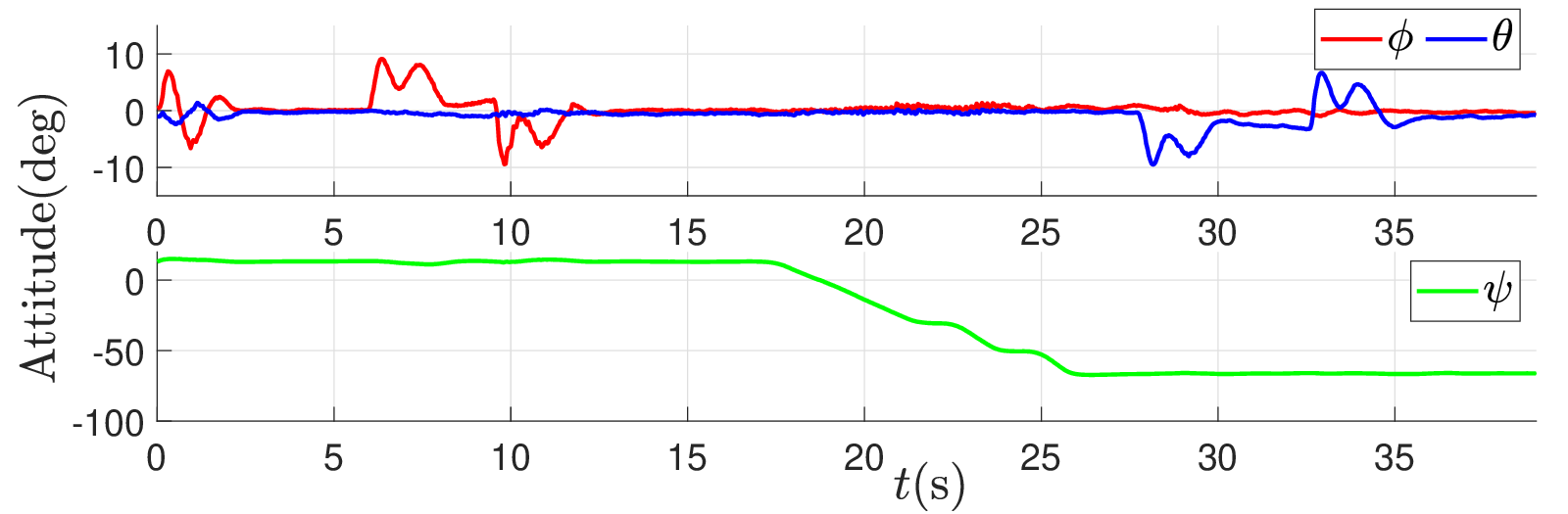}}
\subfigure[tension force estimation]
{\label{figure:experiment_flight_Ft_18}\includegraphics[width=0.95\linewidth]{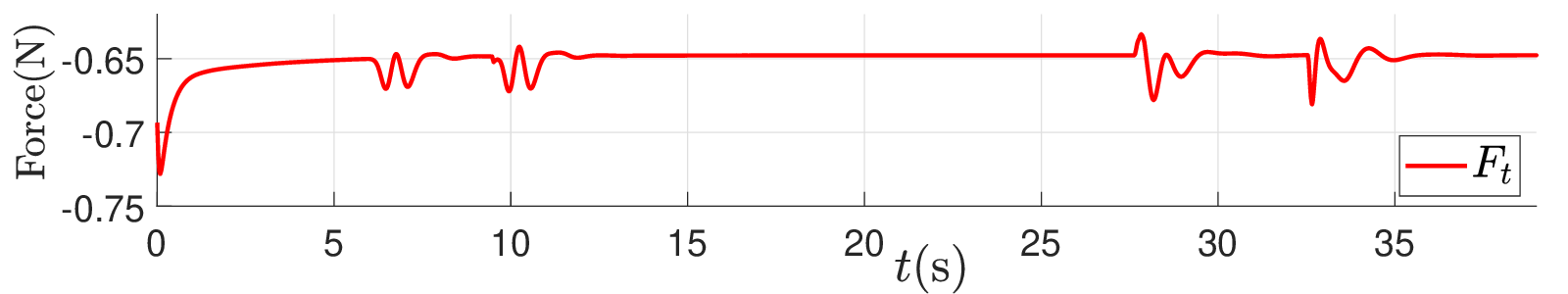}}
\subfigure[Control signals of rotors]
{\label{figure:experiment_flight_pwm_18}\includegraphics[width=0.95\linewidth]{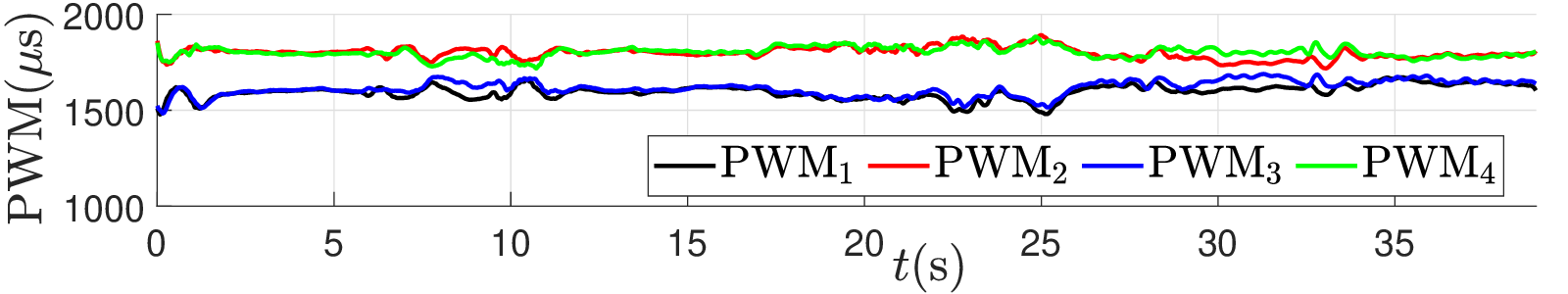}}
\caption{Results of flight experiment ($\bm{L}=[-0.18~0~-0.05]^\top m$).}
\label{figure:experiment_flight_18}
\end{figure}

In the real flight experiment, the control objective is to track the desired load velocity $\bm{\dot{\xi}_p}$. 
Under the BS controller, the UAV fails to maintain stable flight due to significant disturbances induced by the slung load. 
Consequently, we only present the real flight experimental results of the proposed control strategy. 
In addition, considering the structural symmetry of the quadrotor, the offset is introduced only along the axes $x_b$ and $z_b$. 
The results of the real flight experiment with the offsets $\bm{L}=[-0.12~0~-0.05]^\top\text{m}$ and $\bm{L}=[-0.18~0~-0.05]^\top\text{m}$ are presented in Fig. \ref{figure:experiment_flight_12}, \ref{figure:experiment_flight_18}, respectively. 
The performance metrics of the experimental results are presented in Table \ref{tab:experiment_velocity}.

In the flight experiment with offset $\bm{L}=[-0.12~0~-0.05]^\top\text{m}$, the UOSL takes off with its  built-in PID controller of the bare UAV, and a swing motion is imposed on the UOSL. 
Then, from $0\text{s}$, the controller switches to the proposed the controller, which successfully stabilizes the system within $2.5\text{s}$. 
From $8.7\text{s}$ to $13.2\text{s}$, the UOSL tracks the desired velocity $\dot{y}_{pd}=1.5 \text{m/s}$, the load velocity $\dot{y}_{p}$ converges to the range of $[1.35,~1.65]\text{m/s}$ in $3.04\text{s}$, and the overshoot of the step response is $16\%$. The RMSE of $\dot{y}_{pd}$ in this phase is $0.814\text{m/s}$. 
From $44.19\text{s}$ to $49.25\text{s}$, the UOSL tracks the desired velocity $\dot{x}_{pd}=1.5 \text{m/s}$, the load velocity $\dot{x}_{p}$ converges to the range of $[1.35,~1.65]\text{m/s}$ in $2.8\text{s}$, and the overshoot of the step response is $11.3\%$. 
The RMSEs of swing angles $\alpha$ and $\beta$ in the whole flight test are $1.71^\circ$ and $2.4^\circ$, respectively. 
Lastly, the UOSL control is switched to its  built-in  PID controller to complete the landing. 
In the next real flight experiment with the offset $\bm{L}=[-0.18~0~-0.05]^\top\text{m}$ with the results shown in Fig. \ref{figure:experiment_flight_18}, the UOSL successfully achieves similar maneuver by the proposed control law. 
Finally, we conclude that the proposed control strategy can achieve velocity tracking and active anti-swing control for the UOSL with different offsets $\bm{L}$. 
The entire experimental process does not rely on any external positioning system, such as RTK or motion-capture system, and the UOSL obtains its states $\bm{q}$ and $\bm{\dot{q}}$ solely from the onboard IMU, gyroscope, and optical-flow sensors. 
To the best of our knowledge, without relying on any external positioning systems, this is the first real flight experiment on a UOSL system. 
According to Fig. \ref{figure:experiment_flight_Ft} and \ref{figure:experiment_flight_Ft_18}, the proposed control law successfully estimates the tension force $F_t$ generated by (\ref{model_xi1}) acting on the cable. The results show that $F_t$ consistently fluctuates around the gravitational force of the slung load, calculated as $0.066\text{kg} \times g = 0.6472\text{N}$, indicating a reliable tension force estimation throughout the flight process. 
The video of the real flight experiment is available: 
\url{https://youtu.be/tQS3m1oJ-U4}.


\begin{table}[!htb]
\scriptsize 
\centering
\caption{Quantitative analysis for experimental results}
\begin{tabular}{c|ccc|ccc}
\hline\hline
&Settling&Max&RMSE&Settling&Max&RMSE \\
&time(s)&overshoot& $(\text{m/s}~\text{or}~^\circ)$
&time(s)&overshoot& $(\text{m/s}~\text{or}~^\circ)$\\
\hline
&\multicolumn{3}{c}{$\bm{L}=[-0.12~0~-0.05]^\top \text{m}$}&\multicolumn{3}{c}{$\bm{L}=[-0.18~0~-0.05]^\top \text{m}$} \\
\hline
$\dot{x}_p$     &2.8    &11.3\%     &0.732
                &2.22	&20\%	    &0.699\\
$\dot{y}_p$ 	&3.04   &16\%       &0.814
                &2.66   &19.3\%     &0.78\\
$\alpha$ 	&-   &-       &1.71
            &-   &-       &1.55\\
$\beta$ 	&-   &-       &2.4
            &-   &-       &2.14\\
\hline
\hline
\end{tabular}
\label{tab:experiment_velocity}
\end{table}




\section{Conclusions and Future Work}
In this paper, a new dynamic model for the UOSL system is constructed, based on which a nonlinear control is developed. The proposed control scheme is implemented on a UOSL experimental platform, and both simulations and real flight experiments have been conducted to validate its effectiveness. The satisfactory results demonstrate the practicality and robustness of the proposed method. Importantly, this work provides a novel solution for controlling mechanical systems with  built-in  off-center characteristics. 
In the future work, we plan to extend the proposed control framework to more advanced scenarios, including multi-UAV cooperative load transportation. Furthermore, learning based adaptive control can be designed and implemented to further enhance the system's robustness and scalability.



\bibliographystyle{ieeetr}
\bibliography{References}

\begin{IEEEbiography}[{\includegraphics[width=1in,height=1.25in,clip,keepaspectratio]{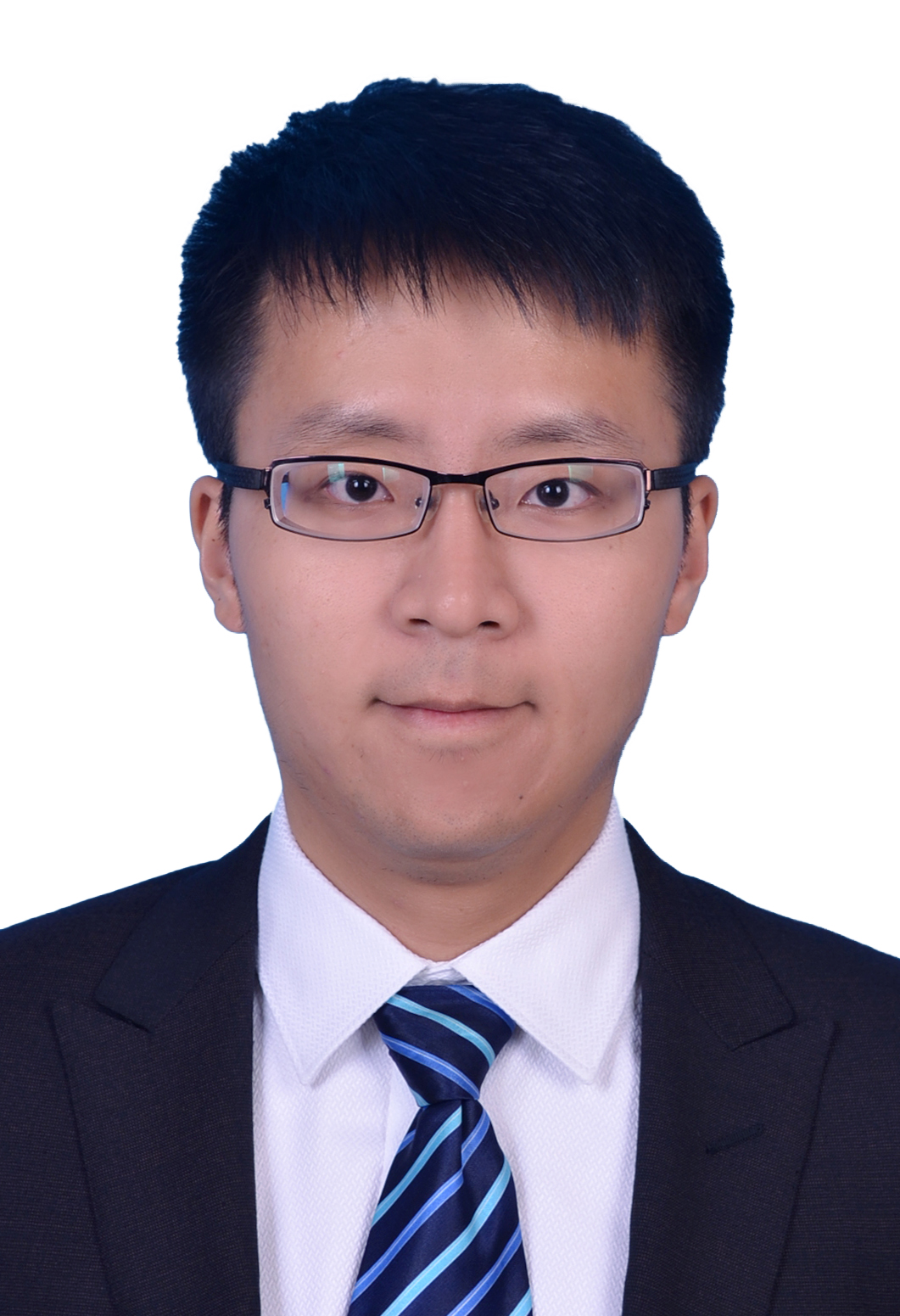}}]{Zongyang Lv}(Member, IEEE)%
received the M.S. degree in mechanical engineering and the Ph.D. degree in control theory and control engineering from the Dalian
University of Technology, Dalian, China, in 2015 and 2021, respectively. 

From 2022 to 2023, he was a postdoctoral researcher with the Dalian University of Technology, Dalian, China. 
He is currently a postdoctoral researcher with the Department of Electrical and Computer Engineering, University of Alberta, Edmonton, Canada. 
He is the winner of the Best Paper Prize Award of IFAC Control Engineering Practice (2023). His research interests include nonlinear control, finte-time control, and unmanned aerial vehicle control.  
\end{IEEEbiography}
\begin{IEEEbiography}[{\includegraphics[width=1in,height=1.25in,clip,keepaspectratio]{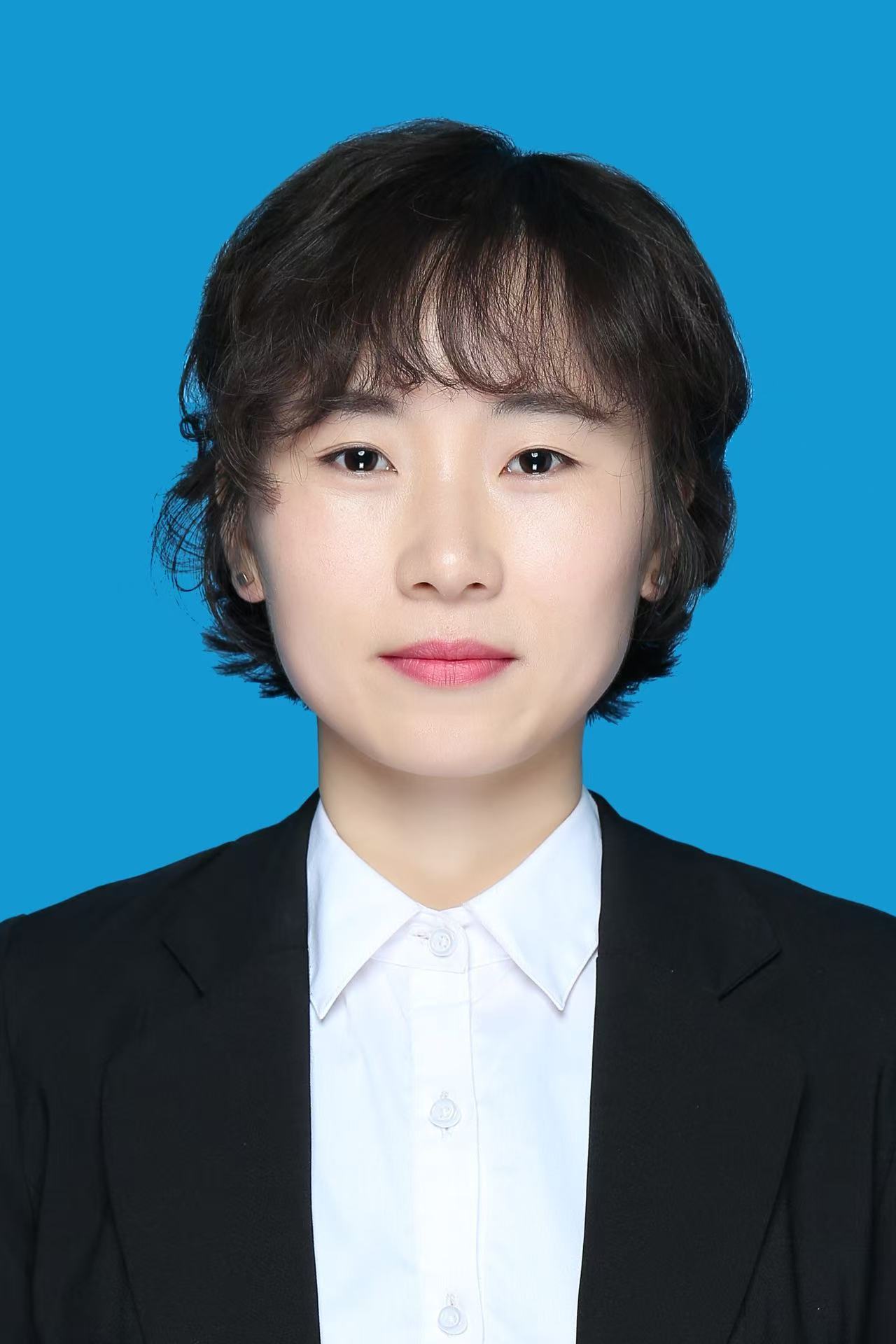}}]{Yanmei Jia}%
received the Ph.D. degree in mathematics from the Dalian University of Technology, Dalian, China, in 2020. She joined the School of Science, Dalian Minzu University, Dalian, China, as a Lecturer in 2021. Her current research interests include optimization and nonlinear control theory and control applications in unmanned aerial vehicles. 
\end{IEEEbiography}
\begin{IEEEbiography}[{\includegraphics[width=1in,height=1.25in,clip,keepaspectratio]{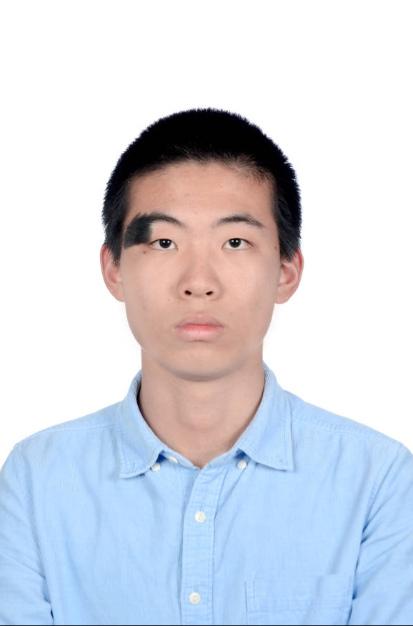}}]{Yongqing Liu}
received the B.Sc. degree in Electrical Engineering from the University of Alberta, Edmonton, Canada, in 2023. He worked as a research assistant at the Faculty of Electrical and Computer Engineering, University of Alberta, during the summers of 2023 and 2024. He is currently pursuing the M.Eng. degree at the University of Alberta, where his research focuses on control systems for UAVs.
\end{IEEEbiography}
\begin{IEEEbiography}[{\includegraphics[width=1in,height=1.25in,clip,keepaspectratio]{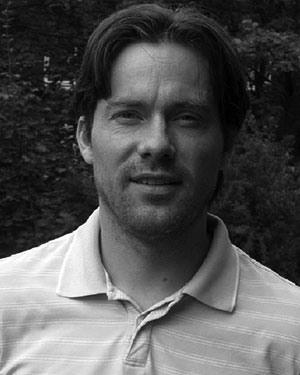}}]{Alan Lynch}(Member, IEEE) received the B.A.Sc. degree in engineering science (electrical option) from the University of Toronto, Toronto, ON, Canada, in 1991; the M.A.Sc. degree in electrical engineering from the University of British Columbia, Vancouver, BC, Canada, in 1994; and the Ph.D. degree in electrical and computer engineering from the University of Toronto, in 1999. 

Since 2001, he has been a Faculty Member with the Department of Electrical and Computer Engineering, University of Alberta, Edmonton, AB, Canada, where he is currently a Full Professor. His interests include nonlinear control and its applications to electrical and electromechanical systems including power converters, unmanned aerial vehicles, and self-bearing motors.
\end{IEEEbiography}
\begin{IEEEbiography}[{\includegraphics[width=1in,height=1.25in,clip,keepaspectratio]{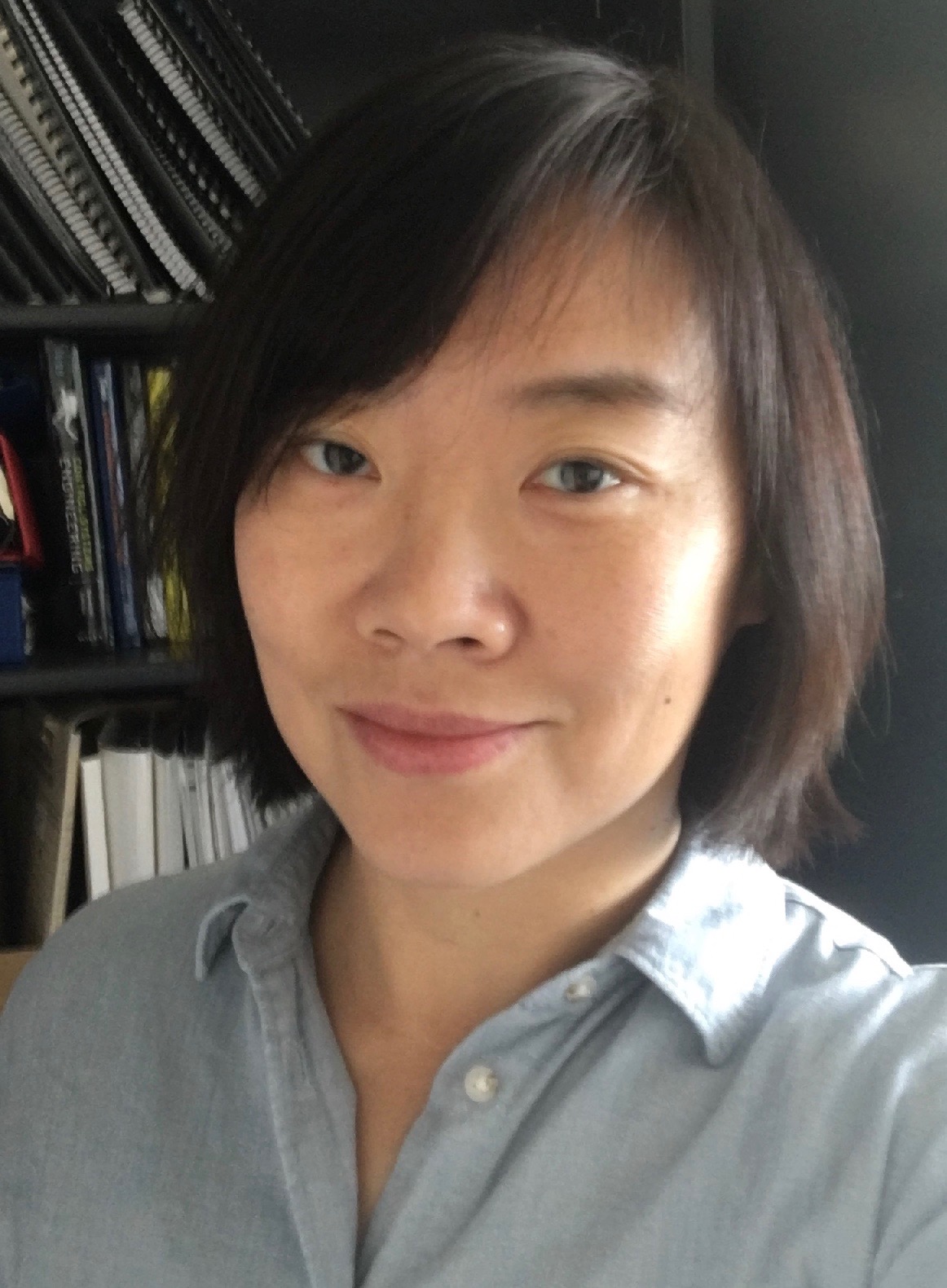}}]{Qing Zhao}(Member, IEEE)
received the B.Sc. degree in Control Engineering from Northeastern University (NEU), China, and received the Ph.D. degree in Electrical Engineering from the Western University (formerly University of Western Ontario), London, Ontario, Canada. 
She is currently a Professor in the Department of Electrical and Computer Engineering at the University of Alberta, Edmonton, Alberta, Canada. She received the A. V. Humboldt Research Fellowship for Experienced Researchers in 2009, while she was on sabbatical leave in Germany and Belgium. She is a registered professional engineer (PEng) in Alberta, Canada. Her research interests include fault diagnosis, fault tolerant control, machine condition monitoring, and industrial data analytics.
\end{IEEEbiography}
	
\begin{IEEEbiography}[{\includegraphics[width=1in,height=1.25in,clip,keepaspectratio]{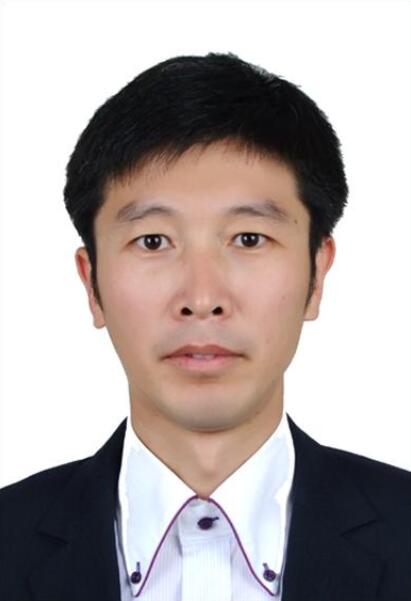}}]{Yuhu Wu}
		(M'15) received the Ph.D. degree in mathematics from the Harbin Institute of Technology, Harbin, China, in 2012.
	
		Since 2012, he has held an Assistant Professor position with the Harbin University of Science and Technology, Harbin. He held a Postdoctoral Research position with Sophia University, Tokyo, Japan, from 2012 to 2015. In 2015, he joined the School of Control Science and Engineering, Dalian University of Technology, Dalian, China, where he is currently a Full Professor. His research interests are related to optimization, and nonlinear control theory and applications of control to Boolean networks, automotive powertrain systems, and unmanned aerial vehicles.
	\end{IEEEbiography}
\end{document}